\documentclass[article]{aa} 
\usepackage{graphicx}
\usepackage{txfonts}
\usepackage[colorlinks=true,citecolor=blue]{hyperref}
\usepackage{graphicx}
\usepackage{amsmath,amssymb}
\usepackage{times}
\usepackage{natbib}
\usepackage{color}
\usepackage{multirow}
\usepackage{todonotes}
\usepackage{appendix}
\usepackage{array}
\usepackage{verbatim}
\usepackage{enumitem}
\usepackage{pifont}

\definecolor{DarkBlue}{rgb}{0.1,0.2,0.7} % blu scuro
\newcommand{\msol}{\rm M_{\odot}}

\newcommand{\kpc}{\mathrm{kpc}}
\newcommand{\Gyr}{\mathrm{Gyr}}
\newcommand{\kms}{\mathrm{km/s}}

\begin{document} 

\title{Cold gas formation triggered by active galactic nuclei jet feedback in galaxy cluster cores} 

\titlerunning{Jet-triggered cold gas formation}
\authorrunning{S. Sotira et al.}

\author{S.  Sotira\inst{1,2}\fnmsep\thanks{\email{stefano.sotira@unibo.it}}, M. A. Bourne\inst{3,4}, D. Sijacki\inst{4,5}, F. Vazza\inst{1,6} and F. Brighenti\inst{1,7} }
\institute{Dipartimento di Fisica e Astronomia, Università di Bologna, Via Gobetti 93/2, 40129 Bologna, Italy
\and
INAF - Osservatorio di Astrofisica e Scienza dello Spazio di Bologna, Via Gobetti 93/3, 40129 Bologna, Italy
        \and  Centre for Astrophysics Research, Department of Physics, Astronomy and Mathematics, University of Hertfordshire, College Lane, Hatfield AL10 9AB, UK
        \and
         Kavli Institute for Cosmology, Cambridge, University of Cambridge, Madingley Road, Cambridge CB3 0HA, UK
         \and  Institute of Astronomy, University of Cambridge, Madingley Road, Cambridge CB3 0HA, UK
         \and
            Istituto di Radio Astronomia, INAF, Via Gobetti 101, 40129 Bologna, Italy  
            \and
            University of California Observatories/Lick Observatory, Department of Astronomy and Astrophysics, Santa Cruz, CA 95064, USA
        }

   \date{Received / Accepted}

% \abstract{}{}{}{}{} 
% 5 {} token are mandatory

\abstract    
{
%introduction
Extended warm and cold gas nebulae, with complex morphologies and kinematics, have been observed in the centres of cool-core galaxy clusters. Their origin within the hot intracluster medium (ICM) is still puzzling, and among many mechanisms, positive feedback from the central active galactic nucleus (AGN) has been proposed.
% aim and method
In this work, we performed a suite of very high-resolution hydrodynamic simulations of a Perseus-like cool-core galaxy cluster subject to self-regulated AGN jet feedback, which leads to realistic ICM properties. By explicitly following warm ionized, neutral, and molecular gas phases, we studied the complex interplay between AGN activity and the multi-phase ICM.
%result
While AGN feedback globally heats the ICM, we find that during the individual AGN jet bursts, hot material is also injected laterally to the jet axis, within the turbulent mixing layer. This material, as it expands, compresses the surrounding hot ICM, reducing the local cooling time, and leads to the formation of cold clumps on a characteristic timescale of $\sim 30$~Myr. By employing tracers, we explicitly track cooling within the affected regions, finding that very hot gas identified in high-compression, low-vorticity zones condenses in situ to form cold clumps. A statistical analysis reveals that the condensation of cold gas is highly promoted once the local turbulent Mach number, $\sigma_{\mathrm{hot}} / c_{s, \mathrm{hot}}$, in the hot gas component ($T \geq 10^7\ \mathrm{K}$) takes values around $ \sim 0.3$.
%conclusion
The presented process is a further important step in understanding the physical mechanisms that lead to the formation of cold gas in the cluster core. Our measured values of the characteristic turbulent Mach number, together with detailed multi-phase gas kinematics predictions, provide important theoretical tools to interpret future X-ray spectroscopy and deep radio data, ultimately to constrain the origin of cool-core cluster nebulae. 
}

\keywords{galaxy clusters, general --
             methods: numerical -- 
             intergalactic medium -- turbulence -- %galaxies:cluster:intracluster medium -- 
             galaxies: jets
             }

\maketitle

\section{Introduction} \label{sec:introduction}
Within cool-core galaxy clusters, the feedback from the central active galactic nuclei (AGN) has been firmly identified as the fundamental process regulating intracluster medium (ICM) thermodynamic properties through providing sufficient energy supply to combat gas radiative cooling losses \citep[see e.g.][]{2007ARA&A..45..117M, Fabian12, Hlavacek-Larrondo2022, BourneYang23}.

However, surrounding the cluster-Dominant (cD) galaxy, large, clumpy, filamentary, and multi-phase gas structures have been widely observed. The warm ionized phase ($T\simeq10^4\, \mathrm{K}$) has been detected through $\mathrm{H\alpha}$, [N II], [O II], OI and [Si II] emission lines, revealing structures with highly variable sizes that range from a few kpc up to $\sim80-100\ \kpc$ \citep{Fabian08, Fabian16, Fabian12, Olivares2019, olivares22, Gingras24, Tamhane2026}. Co-spatial with these structures is a molecular component observed in CO and H2 emission \citep{Salome11, Pulido18, olivares22}.
The mass of the molecular gas in these filaments ranges between $10^8-10^{10}\ \msol$, and the mass of the ionized gas is of the same order of magnitude. Furthermore, these filaments also contain dust and are permeated by magnetic fields \citep{Fabian08, Donahue2011, Reefe2025}. Previous work has suggested that the magnetic fields may stabilize the filaments against gravitational collapse \citep{Fabian08}, but some observations also reveal the presence of star formation \citep{Fogarty15, Fogarty19}.

Observations of cold gas kinematics reveal line-of-sight velocities ranging from $\pm50$ to $\pm600\ \kms$ \citep{Gendron-marsolais18, Olivares2019, olivares22,Ganguly23, Gingras24,Li25}. The complex morphologies and kinematics allow, in a few cases, the determination of whether the gas is infalling or outflowing with respect to the central supermassive black hole (SMBH). The main caveats here are projection effects and the determination of the filling factor. Understanding the detailed kinematics of the hot phase is, however, more challenging. So far, there are only a handful of measurements of the hot ICM velocity dispersion using X-ray spectroscopy, and they have a lower spatial and spectral resolution compared to the cold phase measurements. Somewhat surprisingly perhaps, these observations have revealed relatively quiet environments for the few cluster cores observed so far, with velocity dispersion values between $150-200\ \kms$ \citep{hitomi16, Fujita25, XRISMColl25may, XrismCol25March}.

Given this complex observational picture, the nature and formation of these cold filaments are still not well understood, notwithstanding a large theoretical body of work. Linear stability analysis \citep[e.g., ][]{Malagoli87, Balbus1989, Loewenstein1999} showed that in a stratified galaxy cluster atmosphere, thermal instability cannot form localized clouds of cold gas too distant from the halo centre: perturbations are either over-stable or their cooling time is larger than their free-fall time.

However, the evolution of non-linear perturbations, generated by AGN feedback, for instance, seems more relevant in this context. It is plausible that a strong enough perturbation could rapidly cool and form a cold clump \citep[e.g., ][]{Joung2012, Beckmann19}, with the density (or entropy) contrast threshold for localized cooling being determined both by the properties of the background medium and the nature of the perturbation. \citet{Voit2017} summarized the basic underlying theoretical understanding of the cold gas cycle in the ICM subject to AGN feedback. As the gas is uplifted by the central AGN activity, new cold gas may form through thermal instability at larger radii caused by turbulence and mixing of the uplifted low-entropy gas. Several groups suggested that this happens when the minimum of the ratio between the cooling time, $t_{\mathrm{cool}}$, and the free-fall time, $t_{\mathrm{ff}}$ in the cluster is approximately equal to or below 10 \citep[e.g., ][]{gaspari12, mccourt12, VoitDonahue2015,ChoudhurySharma2016}, even though cold gas is also found in simulations with different values of this ratio, and it is not entirely clear whether this captures the whole physics involved in this process \citep{McNamara16, Voit2017, Beckmann19}. 
For example, higher values of the ratio $t_{\mathrm{cool}}/t_{\mathrm{ff}}$ can still lead to the condensation of cold gas if large density perturbations are present, as shown by \citet{Choudhury2019} using 2D hydrodynamical simulations, or if there is enough turbulence in the medium to make buoyancy oscillations thermally unstable \citep{Voit2018}. Furthermore, 
\citet{Voit2021} studied cold gas condensation and precipitation due to different astrophysical processes in cluster and galactic halos, generalizing previous findings. Entropy fluctuations, which in turn are directly linked to density and temperature perturbations \citep[see also e.g.,][]{Fielding2020},  and the tails of the respective distributions, have been found to be the key in driving the cold gas precipitation from a hot atmosphere.

A large body of theoretical work, studying cold gas formation due to AGN jet or cavity propagation, found that the complex multiphase medium can extend to tens of kpc from the cluster centre \citep[e.g.,][]{Brighenti2002, Revaz08, gaspari12, Gaspari13, Gaspari18, LiCC2014, Brigh15, Beckmann19, Qiu20, Fournier24, Sotira25}. Part of this cold gas is expected to infall towards the innermost regions, accrete onto the central SMBH on a dynamical time, and hence sustain the AGN feedback cycle. For example, \citet{Revaz08} used an idealized simulation setup to show how cold gas can condense at the edges of rising cavities, falling back towards the centre of the gravitational potential. \citet{Brigh15}, using 2-dimensional simulations showed that cold clumps can form in the wake of a rising cavity and suggested that large vortexes generated around the buoyantly rising cavities lead to sustained, long-lasting compression. 
\citet{Rhea25} studied the impact of magnetic fields on the survival of cold filaments, which can also help to promote cold clump filamentary morphology \citep{Wang2021, Ehlert23, DasGronke24, Fournier24}. Furthermore, \citet{LiCC2014}, \citet{Beckmann19} and \citet{Fournier24} conducted a detailed analysis of the properties of the cold gas substructures in AGN feedback simulations and found that the AGN activity has a positive impact on the cold gas formation, with an important fraction of the cold gas forming during the uplift or condensing during the fallback toward the centre. \citet{Jennings2023} also found that the filaments are likely to experience a natural fragmentation in the hot ICM and that the resulting small clumps are sites for further condensation.

Investigating these phenomena with numerical simulations has, however, many limitations. The relevant temperature range is very large, since the ICM contains multi-phase gas from the X-ray emitting hot plasma at $\sim 10^8\ \mathrm{K}$ to the very cold molecular phase at $10-100\ \mathrm{K}$. Furthermore, when reaching the very cold phases of the gas, it is necessary to take care of the numerous (non-equilibrium) thermochemical processes happening at these low temperatures, as well as the complex physics of star formation and stellar feedback. 
To circumvent these difficulties, numerical simulations of galaxy clusters often introduce an artificial gas temperature floor of $10^4\ \mathrm{K}$, with the obvious drawback that this does not allow a complete view of the formation of the cold clumps \citep[e.g., ][]{LiCC2014}, and also likely limits our understanding of AGN fuelling. 

Another critical point is the AGN feedback implementation: even high-resolution zoom-in simulations of galaxy clusters cannot inject AGN-driven jets from first principles and instead have to employ subgrid models \citep[for a review see][]{BourneYang23}. 
While some works compute the AGN power from the accretion of the hot gas based on the Bondi-like formula \citep[e.g., ][]{Sijacki2007, dubois11}, in other works the AGN feedback power is calculated taking into account the quantity of gas below a temperature threshold (usually a few $10^4\ \mathrm{K}$) or above a given density threshold, within a region close to the sphere of influence of the SMBH \citep[e.g., ][]{gaspari11b, gaspari11a, Gaspari2015, Libryan14b, LiCC2014, Prasad2015, YangRey16, Ehlert23}.

In our first work, \citet{Sotira25}, through numerical simulations based on the approach of \citet{Libryan14b} and adopting the \textsc{Enzo} code, we found that the hot gas velocity dispersion in the cluster core is strongly linked to the AGN activity and that, even though to first order the kinematics of the hot and the colder phases are roughly correlated, they can show large discrepancies when analyzed in detail. Furthermore, we found that the formation of cold gas is promoted by kinetic jet feedback and further aided by jet precession. However, a fully self-consistent model that explains how the colder phases can be formed in the hot environment of the cluster core is still lacking, and is the aim of the present study.

Here, using the moving-mesh code \textsc{Arepo} \citep{Springel2010, Pakmor2016}, we perform a new simulation suite following the AGN feedback cycle in a Perseus-like galaxy cluster. We use the {\sc Grackle} library \citep{Smith16}, which models non-equilibrium primordial chemistry, as well as metal cooling via look-up tables, and allows for low temperature cooling to follow the formation of the warm ionized, neutral, and molecular phases, with a minimum temperature threshold at $10\ \mathrm{K}$. This approach allows us to have more reliable and detailed analysis of the formation and kinematics of the cold, warm ionized and molecular phases, and their connections with the hot medium. We,  furthermore, incorporate jet-driven AGN feedback based on work by \citet{BourneSij17, BournSij19} and \citet{bourneSij2021}, together with a novel sink-particle-based model for SMBH accretion (Ortame et al., in prep), allowing us to simulate very high-resolution hydrodynamical jets and resolve the lobe and ICM turbulence thanks to targeted super-Lagrangian refinement schemes. These high-resolution, light jets do not `drill' through the ICM, but can be deflected by the cold gas phase present in the ICM.  

The paper is organised as follows: in Section~\ref{sec:methods_sim}, we introduce the simulation setup, describing the initial conditions and the feedback implementation; in Section~\ref{sec:results-sanity_check}, we present the global properties of our simulation suite, verifying the validity of the outcomes; in Section~\ref{sec:results_cold-gas}, we discuss the physical model responsible for the cold gas formation. Finally, in Section~\ref{sec:conclusions}, we discuss the critical aspects of the model and present our conclusions.

\section{Methodology and Simulation Suite}  \label{sec:methods_sim}

\subsection{Numerical code and initial conditions}
Simulations presented here were performed using the moving-mesh code \textsc{Arepo} \citep{Springel2010, Pakmor2016}. Our simulated galaxy cluster has physical properties closely resembling the Perseus cluster. The initial gas density profile is taken from the X-ray fitting of \cite{Churazov04},
\begin{equation}    
n(r) =\left( \frac{0.046}{\left[1+\left(\frac{r_{\mathrm{kpc}}}{57}\right)^2\right]^{1.8}}+
\frac{0.00479}{\left[1+\left(\frac{r_{\mathrm{kpc}}}{200}\right)^2\right]^{0.87}}\right) \ \ \ \ \ \ \mathrm{cm}^{-3}.
\end{equation}
We assume an analytic dark matter halo that follows the NFW distribution with a mass profile 
\begin{equation}
M_{\mathrm{h}}(r)=\frac{4\pi}{3} \rho_cR_s^3d_c\left(\log\left(1+r/R_s\right)-\frac{r/R_s}{1+r/R_s}\right),
\end{equation}
and density profile
\begin{equation}
\rho_{\mathrm{h}}(r)= \rho_c \frac{d_c}{\frac{r}{R_s}(1+r/R_s)^2},
\end{equation}
where $c=5$, $R_s = R_{200}/c$, $R_{200}=1.9\ \mathrm{Mpc}$, $\rho_c=3H_0^2/8\pi G$, and $d_c=200c^3/\big(\log(1+c)-c/(1+c)\big)$.
The gas metallicity profile follows the observational fit by \cite{Rebusco05},
\begin{equation}
    Z(r)=0.3\frac{2.2+(r/80)^3}{1+(r/80)^3}Z_{\odot},
\end{equation}
where $Z_{\odot}=0.127$ is the solar metallicity \citep{Asplund2009}.

The whole simulated domain is a box with side $6R_{200}=11.5\ \mathrm{Mpc}$. We set the minimum allowed volume, $V_{\mathrm{min}}$, for a cell to be the same as a sphere of radius $100\ \mathrm{pc}$.
The target mass of the gas cells is $m^{\mathrm{target}}_{\mathrm{cell}}=10^8\  \msol$, although we note that, due to novel refinement schemes (discussed below), much higher mass resolution is achieved in and around the jets and SMBH regions, reaching minimum cell masses around $10^2-10^3\ \msol$.
The SMBH has an initial mass $M_{\mathrm{BH}}=4\times10^9\ \msol$, and is placed at a fixed position at the centre of the cluster. At the beginning of the simulation, we evolve the cluster adiabatically for $2\ \Gyr$ to relax the initial conditions and to allow temperature and entropy to settle into a hydrostatic equilibrium. 
This procedure also introduces to the ICM a low-level turbulence with velocities around $10\ \kms$.

\subsection{Cooling processes: neutral gas, molecular gas, and dust}
Since the aim of the work is the study of the cold gas phases, we want to follow the formation of low-temperature gas below the typical temperature floor adopted by many numerical approaches \citep[e.g., ][]{Sotira25}. In this work, we focus on the formation channels of neutral and molecular gas, while we will address detailed dust and star formation processes in future work.

We use \textsc{Grackle} \citep{Smith16}, which tracks non-equilibrium primordial chemistry and cooling for nine species of H and He, including H2 formation on dust grains, and has tabulated metal-line cooling rates also in the presence of an ultraviolet background radiation field. We note that, although the simulations reach high spatial resolution, the cooling length and internal structure of cold clouds are not fully resolved \citep[e.g.,][]{ReyEtAl2024}; consequently, the exact partitioning of the cold gas into neutral and molecular components, presented below, should be interpreted as an indicative diagnostic of the model behaviour, as opposed to a strict quantitative prediction.

Detailed modelling of dust evolution, growth and destruction in the ICM is beyond the scope of this paper. However, dust acts as an important catalyst for the formation of molecular hydrogen.  Therefore, we employ a simplified dust model within our simulations. The dust is present only in cells in which more than half of the cell mass (considering both hydrogen and helium) is composed of neutral gas, where a local dust-to-gas ratio is assumed to be $M_{\mathrm{dust}}/M_{\mathrm{gas}}=0.009387$ \citep{Pollack1994}.
This assumption takes into account the initial condition of a dust-poor, hot ICM, due to grain sputtering, and dust growth by accretion in cold, dense gas.
The main cooling process of the hot gas component is Bremsstrahlung cooling for all ionized species \citep{Black1981}.

\subsection{Star formation prescription}
Detailed modelling of star formation and stellar feedback is beyond the scope of this study. 
However, to prevent excessive dense gas formation, it is advantageous both from a physical and numerical viewpoint to convert the densest gas cells, which reside in the interstellar medium (ISM) of the cD galaxy, into stars. At the mass and spatial resolutions that can be achieved even in high-resolution isolated galaxy cluster simulations, it is extremely difficult to employ realistic star formation and feedback models, where the multi-phase nature of the ISM is unresolved. 
On the other hand, employing models that treat the ISM by employing an effective equation of state \citep[see e.g.,][]{Springel2003}, also presents difficulties as star formation often proceeds at relatively low gas density thresholds, and gas cooling to low temperatures on the equation of state is often not considered. Therefore, we employ here a different and simple approach, which allows us to hydrodynamically follow cold, dense gas to sufficiently high densities, as follows.
Following \citet{Bourne15}, whenever a gas cell is colder than its Jeans temperature, defined as,
\begin{equation}
T_{\mathrm{J}}=\rho^{1/3}\frac{\mu m_\mathrm{p}G}{\pi k_{\mathrm{B}}}(N_{\mathrm{ngb}}m_{\mathrm{cell}})^{2/3}\ \ \mathrm{K},  \label{eq:jeanst}
\end{equation}
where $\rho$ is the gas cell density, $m_{\mathrm{cell}}$ its mass and $N_{\mathrm{ngb}}=8$, and the cell has a density higher than the threshold density, $n_{\mathrm{SF}}=2\times10^4\ \mathrm{cm}^{-3}$, it is converted into a star particle. No stellar feedback is produced by the star particles, which act as pure N-body particles after their formation. While this is certainly a significant simplification, for the purpose of this work we are concerned that the amount of dense gas formed is realistic in the simulation, so that we can study the interplay between the AGN-driven jet and the multi-phase ICM (see Section~\ref{sec:coldGas_evolution} for details), while we postpone the development of a more sophisticated star formation module to a future work. 
Simulations with AGN feedback and detailed star formation models in galaxy clusters can be found in \citet{LiB15}.

\subsection{SMBH accretion and AGN feedback implementation}

The feedback implementation is based on the model presented in \citet{BourneSij17} and \citet{bourneSij2021} with some modifications to account for interactions between the jet and cold gas and SMBH accretion modelling. Here, SMBH accretion is tracked based on the sink particle model presented in Ortame et al. (in prep). Specifically, for the purpose of this work, it captures the accretion of cold gas within the accretion radius of the SMBH on a fixed accretion timescale \citep[see also e.g.,][]{Libryan14b, LiCC2014, Sotira25}. The gas cells are drained following a mass-weighted Gaussian kernel if the following requirements are satisfied:\\
$\bullet$ They have a temperature below $T_{\mathrm{cold}}=10^4\ \mathrm{K}$;\\
$\bullet$ They are located within the accretion region with radius $r_{\mathrm{acc}}=500\ \mathrm{pc}$ centred on the SMBH;\\
$\bullet$ The sum of the masses of these cells is greater than a minimum mass $M_{\mathrm{acc}}=10^7\ \msol$.\\
The consequent mass that flows toward the SMBH, $\dot{M_{\mathrm{in}}}$, is the total kernel-weighted\footnote{The fraction of mass each eligible cell contributes to $M_{\rm acc}$ is given by $\min \left[ 0.9,\;\exp\left(-4 \frac{r^2}{R_\mathrm{acc}^2} \right)\right]$.} mass of these cells, multiplied by an accretion efficiency, $\eta=0.1$, divided by an accretion time, $\tau_{\mathrm{acc}}=5\ \mathrm{Myr}$, approximately the free-fall time of the accretion region. It is assumed that half of this mass goes into accretion and half is launched in the jet such that $\dot{M}_{\mathrm{acc}}=\dot{M}_\mathrm{J}=\dot{M_{\mathrm{in}}}/2$.

When the AGN feedback is active, inside the accretion region, an ``injection cylinder'', with radius $r_{\mathrm{jet}}$ and height $2h_{\mathrm{jet}}$ is defined centred on the SMBH, in which mass, momentum and energy are injected in to the gas cells in order to launch the jet. The cylinder aspect ratio is fixed such that $r_{\mathrm{jet}}/h_{\mathrm{jet}}=1$. The cylinder target size is determined such that each half cylinder contains a fixed gas mass of $10^4$~M$_{\odot}$ or at least 10 gas cells. An additional criterion is enforced such that the injection region cannot extend beyond the sink boundary. The cylinder and thus the jet axis do not take into account any precession effects, i.e., the jet is always launched along the z-axis.

The feedback implementation employs a hybrid mix of thermal and kinetic energy injection, which adapts the feedback based on its interplay with different gas phases (i.e., dense vs dilute gas). 
Mechanical feedback from hot high-speed jets acts in a different way on the hot ICM and cold, dense clouds, the former being accelerated, while the latter being heated or destroyed. \citet{Klein94a} and more recent works \citep{Armillotta17, GronkeOh18} through analytical studies and idealized hydrodynamic simulations indicate under which conditions a cold gas cloud is destroyed, or can survive and be accelerated, when hit by a rarefied hot wind, even though the survival of cold clouds in a hot wind tunnel in the presence of radiative cooling, magnetic fields and transport processes is still not well understood \citep{Jennings2023}.
For this purpose, we inject kinetic energy in the hot cells, while we inject thermal energy plus momentum in the cold, dense gas cells near the SMBH \citep[see,][for a comparison of kinetic energy versus momentum conserving injection techniques]{BourneSij17}, under the assumption that cold, dense clouds only feel the ram-pressure of the outflow \citep{Bourne2014, Bourne15}. To separate the two feedback coupling behaviours, we use the density as a discriminant, with a density threshold $n_{\mathrm{th}} = 10\  \mathrm{cm}^{-3}$.

\begin{figure*}
\begin{center}
\includegraphics[scale = 0.655]{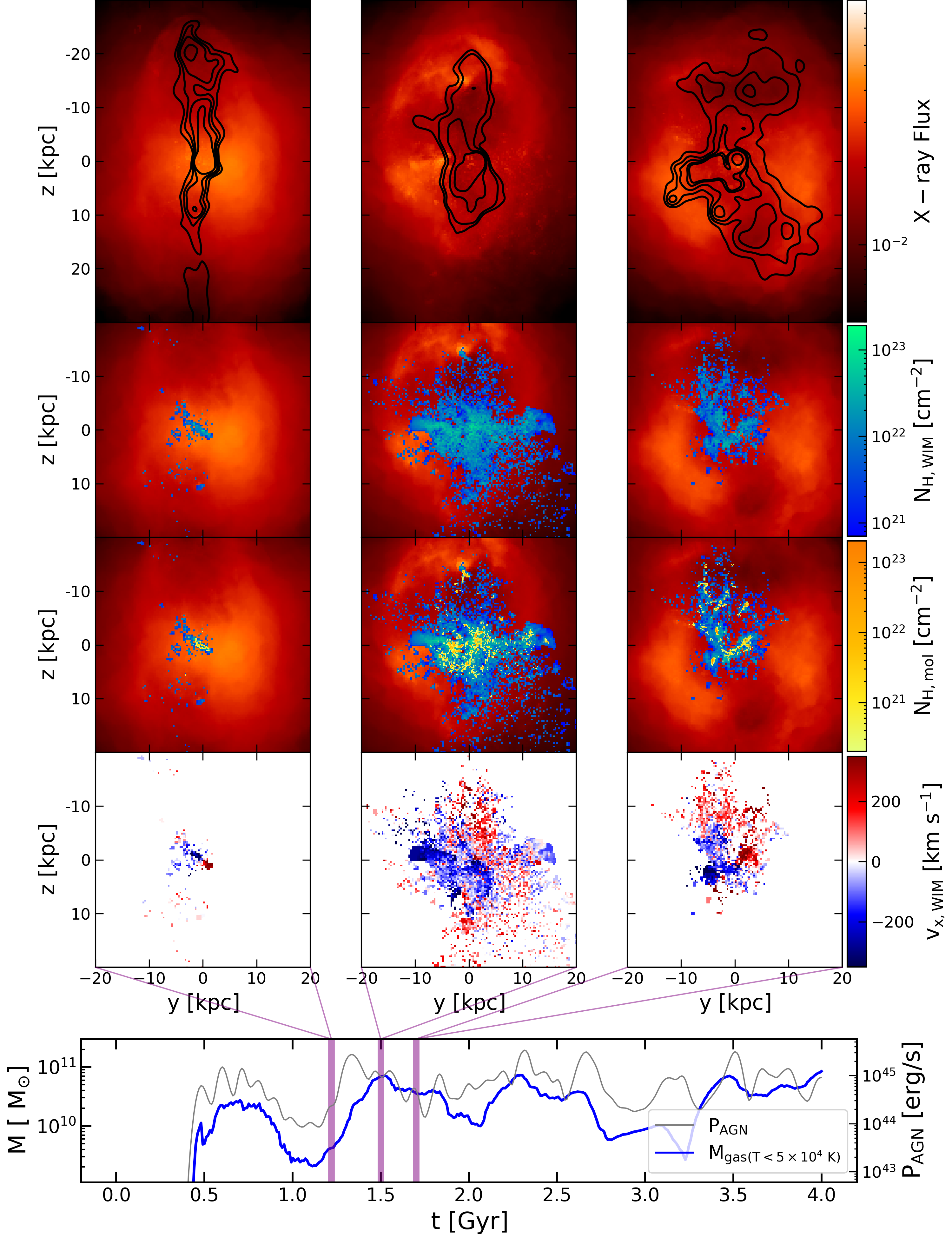}
\caption{Top panels: overview of the gas phases of interest during three particular moments of the simulation (columns from left to right). 
First row: X-ray emission shown with black-red colour tones (see the main text for the emission calculations) and jet tracer projections contours (black colour). X-ray cavities coincide with the locations where jet tracer material is prevalent; second row: Warm ionized gas ($3\times10^3\ \mathrm{K}<T<5\times10^4\ \mathrm{K}$) column density superimposed on the X-ray image, shown with blue-green colour tones; third row: molecular gas column density, superimposed on the previous two images, shown with yellow-orange colour tones.
Fourth row: projected velocity along a line-of-sight of the warm ionized gas phase, which is in good agreement with observations. 
Bottom plot: Total mass of gas with temperature below $5\times10^4\ \mathrm{K}$ (blue curve) and AGN power (grey curve), plotted as a function of time.
Purple bands indicate corresponding time instances of the maps shown. The amount of warm and cold gas correlates with the AGN power because the cold gas, accreting onto the central SMBH, ignites the AGN feedback. The AGN-driven jets propagating through the multi-phase ICM create further perturbations, leading to the renewed formation of cold clumps and filaments. 
}\label{fig:visual}
\end{center}    
\end{figure*}

All quantities are updated in the injection cylinder weighted through a kernel function,
$W_{\mathrm{J}}(r,z)\propto \mathrm{exp}\big(r^2/2r_{\mathrm{jet}}^2\big)|z|$. The expected total mass and energy injected in the $i$-th cell are given by

\begin{equation}
\mathrm{d}E^{\mathrm{tot}}_i=\frac{\dot{E}_{\mathrm{J}}dt}{2}\frac{V_i W_{\rm{J}}(r,z)}{V_{\rm{weight}}},\ \ \ \mathrm{and}\ \ \ 
    \mathrm{d}m_i=\frac{{\dot{M}}_{\mathrm{J}}dt}{2}\frac{V_i W_{\rm{J}}(r,z)}{V_{\rm{weight}}},
    \label{eq:mass_injection}
\end{equation}
respectively, where $\dot{E}_{\mathrm{J}}=\epsilon \dot{M}_{\mathrm{acc}}c^2$ with $\epsilon = 0.1$, $V_{i}$ is the cell volume and $V_{\rm weight}$ is the kernel normalisation. In the kinetic coupling regime, the added momentum to the cell, with initial energy, mass and momentum, $E_{i,0},\ m_{i,0}$, and $ \textbf{\textit{p}}_{i,0}$, respectively, is calculated as
%\begin{equation}
   $|\mathrm{d} \textbf{\textit{p}}^{\mathrm{tot}}_i|=\sqrt{2(m_{i,0}+dm_{i})(E_{i,0}+dE_i^{\mathrm{tot}})}-|\textbf{\textit{p}}_{i,0}|$.
%\end{equation}
This is designed to conserve the injected energy in kinetic form. Due to potential momentum cancellation, if the actual gain in a cell’s total energy is less than  $dE^{\mathrm{tot}}_{i}$, the internal energy will be increased by an appropriate amount to ensure energy conservation.
For the thermal coupling regime, we instead directly inject the assumed jet momentum, with each cell receiving a momentum of
\begin{equation}
\mathrm{d}\textbf{\textit{p}}_i=\frac{\dot{\textbf{\textit{p}}}_\mathrm{J}\mathrm{d}t}{2}\frac{V_i W_{\rm{J}}(r,z)}{V_{\rm{weight}}},
\end{equation}
where $\dot{\textbf{\textit{p}}}_\mathrm{J}=\sqrt{2\dot{M}_{\rm J}\dot{E}_{\rm J}}\hat{\textbf{k}}$. This results in an increase in the kinetic energy of
\begin{equation}
    \mathrm{d}E^{\rm{kin}}_i = \frac{(\textbf{\textit{p}}_{i,0} + \rm{d}\textbf{\textit{p}}_i)^2}{2(m_{i,0} + dm_i)} - \frac{\textbf{\textit{p}}_{i,0}^2}{2m_{i,0}}.
    \label{eq:kinetic_energy_variation}
\end{equation}
Due to the effect of mass loading and, therefore, to conserve the total injected energy, each cell receives a thermal energy injection of $
    \mathrm{d}E_{\mathrm{therm},i} = \mathrm{d}E_{\mathrm{tot},i} - \mathrm{d}E_{\mathrm{kin},i}
    $.

We use two tracers to track jet material. 
The first one, $f_\mathrm{J}$, is set equal to $1$ whenever a cell is in the jet cylinder, and the jet is active. It is subsequently advected with the gas and decays exponentially with time with a decay time of $3\ \mathrm{Myr}$. The primary purpose of this tracer is for the jet refinement scheme (see below). The second tracer, $f_{\mathrm{JM}}$, tracks the mass injected in each cell that belongs to the jet (Eq.~\ref{eq:mass_injection}{\color{red}{b}}), which, like the former tracer, is advected with the gas. This tracer does not decay with time and is used for our analysis (Section~\ref{sec:cold_gas_formation}).

In addition to the standard \textsc{Arepo} refinement criteria based on maintaining an approximately constant gas cell mass, as in \citet{bourneSij2021}, additional criteria are implemented to ensure the jet injection region and jet material maintain high resolution. Specifically, we adopt the super-Lagrangian refinement schemes of \citet{CurtisSijacki2015} and \citet{bourneSij2021} to resolve the region around the SMBH. 
These schemes ensure that within both the accretion region and the jet cylinder, there is a sufficiently high resolution, while maintaining a smooth transition between high and low resolution regions. To ensure that the jets and lobes maintain high resolution, we apply additional jet tracer and property gradient-based refinement criteria similar to those described in \citet{BournSij19} and \citet{bourneSij2021}, that cap the volume of jet/lobe cells based on their tracer fraction, $f_{\rm J}$.

\begin{figure*}
\begin{center}
\includegraphics[scale =0.4]{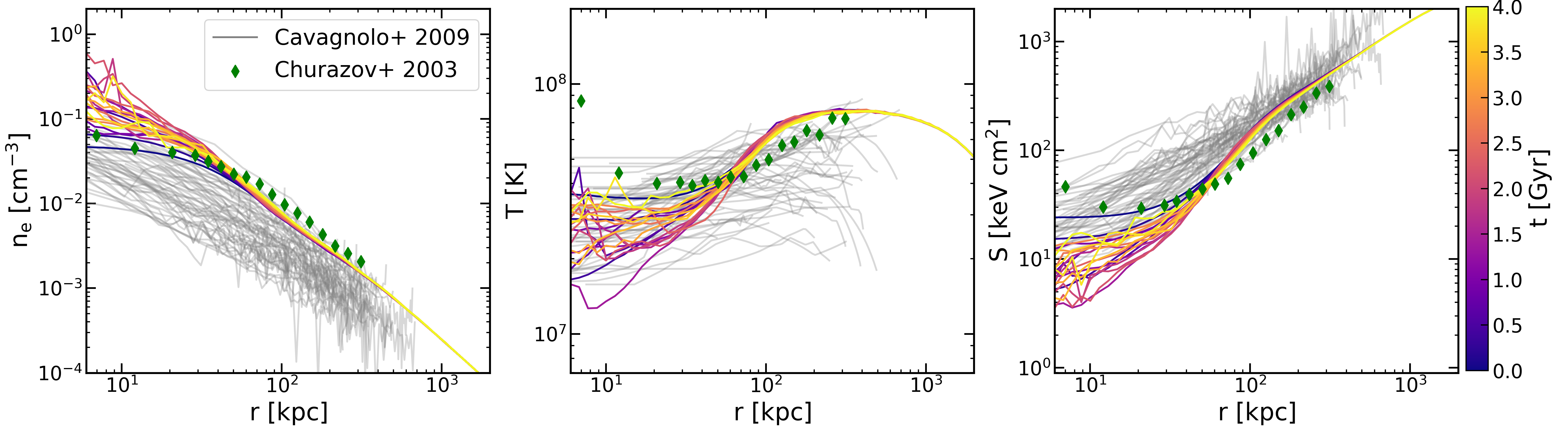}
\caption{From left to the right, X-ray emission-weighted profiles of electron number density, temperature and entropy of our simulations, colour coded with respect to time (purple-yellow colour bar), and compare with: the Perseus cluster derived by \citet{Churazov04} (green diamonds); cool-core clusters from the ACCEPT sample by \citet{Cavagnolo_2008} (grey lines), with a central temperature in a range of $1.5\times10^7\ \mathrm{K} \leq T \leq
5\times10^7\ \mathrm{K}$ and a central entropy less than $50\ \mathrm{keVcm^{-2}}$, i.e. having similar properties to the Perseus cluster. We find that our simulated galaxy cluster profiles are in good agreement with the observed Perseus-like cool-core clusters.}
\label{fig:profiles}
\end{center}    
\end{figure*}

\section{Results : overview of simulations} \label{sec:results-sanity_check}
We begin this section by visualising the multi-phase ICM, outlining the gas kinematics and comparing the ICM radial profiles against the relevant observational sample.

\begin{figure*}
\begin{center}
\includegraphics[scale =0.61]{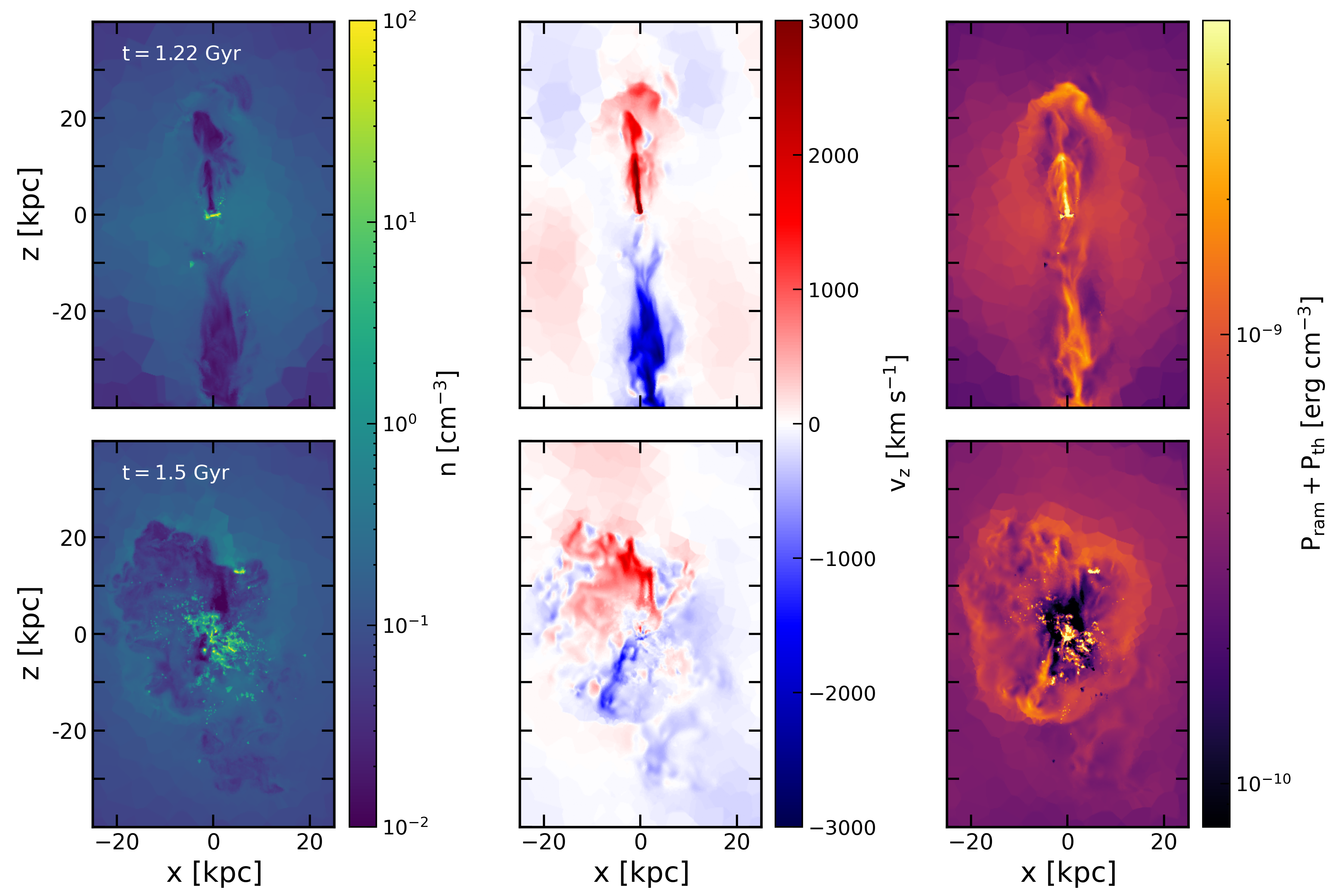}
\caption{ From left to right: maps of the average gas number density, average vertical component of gas velocity, and average total gas pressure (i.e., ram plus thermal) of a $1\ \mathrm{kpc}$ thick projection centred at $y=0$ during the two AGN jet outbursts, at $t=1.22\  \mathrm{Gyr}$ (top) and $t=1.50\  \mathrm{Gyr}$ (bottom). 
With non-relativistic velocities ($v_{\mathrm{jet}}\lesssim 0.1c$), the jets are able to compress the ICM, heating the cluster core and leading to the formation of localized overdensities (see Section~\ref{sec:cold_gas_formation}).} 
\label{fig:dens_vel_slices}
\end{center}    
\end{figure*}

\subsection{Visuals of the multi-phase ICM}
In Fig.~\ref{fig:visual}, we show three representative simulation snapshots (columns from left to right) at different AGN powers and warm and cold gas masses present.
The bottom panel shows the total amount of gas with temperatures below $5\times 10^4\  \mathrm{K}$ and the AGN power (smoothed using a Gaussian filter with a standard deviation of $20\ \mathrm{Myr}$), as a function of time. The times at which panels in each column are produced are indicated by the purple bands.  
The first row shows the X-ray surface brightness (black-red colour tones), computed as $\Sigma_X (x,z) = \int n_{\mathrm H}^2\Lambda_X(T) dy$, where $n_{\mathrm H}$ is the hydrogen number density of the ionized gas and $\Lambda_X(T)$ is the free-free function in the X-ray wavelengths, having non-zero values only in the temperature range $10^6 \ \mathrm{K}\leq T\leq 3\times10^8\ \mathrm{K}$. Projected jet tracer surface densities are shown by overlaid black contours, which are smoothed using a Gaussian filter with a standard deviation of $350\ \mathrm{pc}$ to reduce noise.

The three time instances are taken, respectively, 1) in a period of low AGN activity preceding an AGN outburst; 2) soon after a strong AGN outburst and 3) in a period of mid activity after the AGN outburst. The X-ray maps reveal large cavities which are filled with the jet lobe material, with larger cavities corresponding to higher AGN activity.
At these three different times, we can find, respectively: 1) a relatively low amount of warm and cold gas mostly confined to the centre, 2) a large quantity of warm and cold gas spread throughout the central $30\ \kpc$, and 3) warm and cold gas less spread-out and instead more distributed around the cavities.
In general, significant fractions of the warm-ionized gas (second row) are located in the cluster centre, although this gas component extends throughout all of the core, reaching cluster-centric distances of around $40\ \kpc$. 
Warm-ionized gas has both filamentary and clumpy appearance with some of the gas tracing cavity rims. 
The molecular gas is mostly confined within the inner $10-15\ \kpc$, but it can also be found in smaller amounts within warm clumps. 
This is in line with observations of warm and molecular gas in cluster centres \citep{Salome11, Olivares2019}.

\subsection{Warm-ionized gas kinematics}\label{Sec:kinematics}
While a detailed analysis of the kinematics of the multi-phase gas will be tackled in an upcoming work (Sotira et al., in prep), we give here a cursory exploration of the complexity of gas flows. 
The fourth row of Fig.~\ref{fig:visual} shows the emission-weighted projected velocity along a line-of-sight perpendicular to the jet axis for the warm-ionized gas component. 
Our simulated gas velocities, in the range $\pm50$ -$\pm300\ \kms$, are comparable to those measured in observed clusters \citep{Gendron-marsolais18, Rhea25}. Specifically, \citet{Gendron-marsolais18} studied the kinematics of this phase in the Perseus cluster (see their Figure 5), whereby the simulation results are in very good agreement with their findings, matching both the velocity range and kinematic map topology.

\subsection{ICM radial profiles}
In Fig.~\ref{fig:profiles}, we show the profiles of X-ray emission-weighted electron number density, temperature and entropy, from the left to the right, respectively, colour coded to mark different time instances in the simulation, and compared to the observed profiles for the Perseus cluster \citep[][]{Churazov04}, as well as for selected cool-core clusters from the ACCEPT sample \citep[][]{Cavagnolo_2008}, which includes clusters with similar properties to the Perseus cluster.
Note that the observed Perseus cluster profiles, used to create our initial conditions, are towards the edge of the distribution of cool-core clusters in the ACCEPT sample. Across the entire $4\ \Gyr$ span of our simulation, the simulated cluster does not experience any excessive cooling flow, while due to the self-regulatory AGN feedback, it maintains the characteristic cool-core features, as can be seen from the temperature and entropy profiles.

Around the epoch of $t=2\ \Gyr$, however, in the central $20\ \kpc$, the core is almost one order of magnitude denser with respect to the observed distribution, resulting in a lower central entropy. In this phase, our system appears as a strong cool-core cluster, although the emission from the denser and colder X-ray emitting phase might be absorbed by the warm ionized gas in the ICM \citep[see, e.g.,][]{Fabian22}. At $t =3\ \Gyr$, the density and entropy profiles settle closer to the Perseus values, and they remain in agreement with observed values until the end of the simulation at $4\ \Gyr$. 

\subsection{Jet density and velocity}
In this section, we illustrate key physical properties of the jet and describe its interaction with the ICM. 
In the left-hand panels of Fig.~\ref{fig:dens_vel_slices}, we show the average gas number density of a $1\ \kpc$ thick projection centred at $y=0$ at two epochs during the jet outbursts in the simulation. 
Close to the SMBH, the jet is less dense than the ICM, having density values around $10^{-2}\ \mathrm{cm}^{-3
}$, with a jet-to-ICM density ratio of around $10^{-2}$. Also, compressed regions with higher density are clearly visible around the jet lobes.

The middle panels of Fig.~\ref{fig:dens_vel_slices} show the corresponding vertical gas velocity maps, $v_z$. The jet material velocities reach up to $\sim 10^4\ \mathrm{km/s}$. At larger radii, when the jet material thermalises and is slowed down, the bulk-outward velocity within the turbulent jet lobes decreases to values around several $\sim 10^2\ \mathrm{km/s}$ to $\sim 10^3\ \mathrm{km/s}$.

With these properties, the jet is capable of compressing the ICM while propagating within it. Indeed, if we consider, at the moment of the injection, $\rho_{\mathrm{jet}}\simeq10^{-26}\ \mathrm{g \ cm^{-3}}$, and $v_{\mathrm{jet}}\simeq10^4\ \kms$, the ram pressure of the jet is $P_{\mathrm{ram}}=\rho v^2\simeq10^{-8}\ \mathrm{erg\ cm^{-3}}$, while during the jet propagation later on, $\rho_{\mathrm{jet}}\simeq10^{-25}\ \mathrm{g \ cm^{-3}}$, and $v_{\mathrm{jet}}\simeq10^3\ \kms$, thus $P_{\mathrm{ram}}=\rho v^2\simeq10^{-9}\ \mathrm{erg\ cm^{-3}}$. These jet ram pressure values are significantly larger that the characteristic thermal pressure of the ICM in the cluster core, $P_\mathrm{ICM}\approx n k_\mathrm{B} T\simeq10^{-10}\ \mathrm{erg\ cm^{-3}}$, with $n\simeq10^{-1}\ \mathrm{cm^{-3}}$ and $T\simeq10^7\ \mathrm{K}$.

The right-hand panels of Fig.~\ref{fig:dens_vel_slices} illustrates this point by showing the total gas pressure composed of ram and thermal pressure. Jets and bow shocks generated by their inflation are clearly visible, highlighting the ICM compression caused by AGN feedback. This results in an overall heating of the ICM  and in a localized formation of overdensities (see Section~\ref{sec:cold_gas_formation}). This will be a crucial point for the remainder of the paper.

\section{Results: properties of warm and cold gas components} \label{sec:results_cold-gas}

\subsection{Warm and cold gas evolution}\label{sec:coldGas_evolution}

\begin{figure}
\begin{center}
\includegraphics[width=\linewidth,height=\textheight,keepaspectratio]{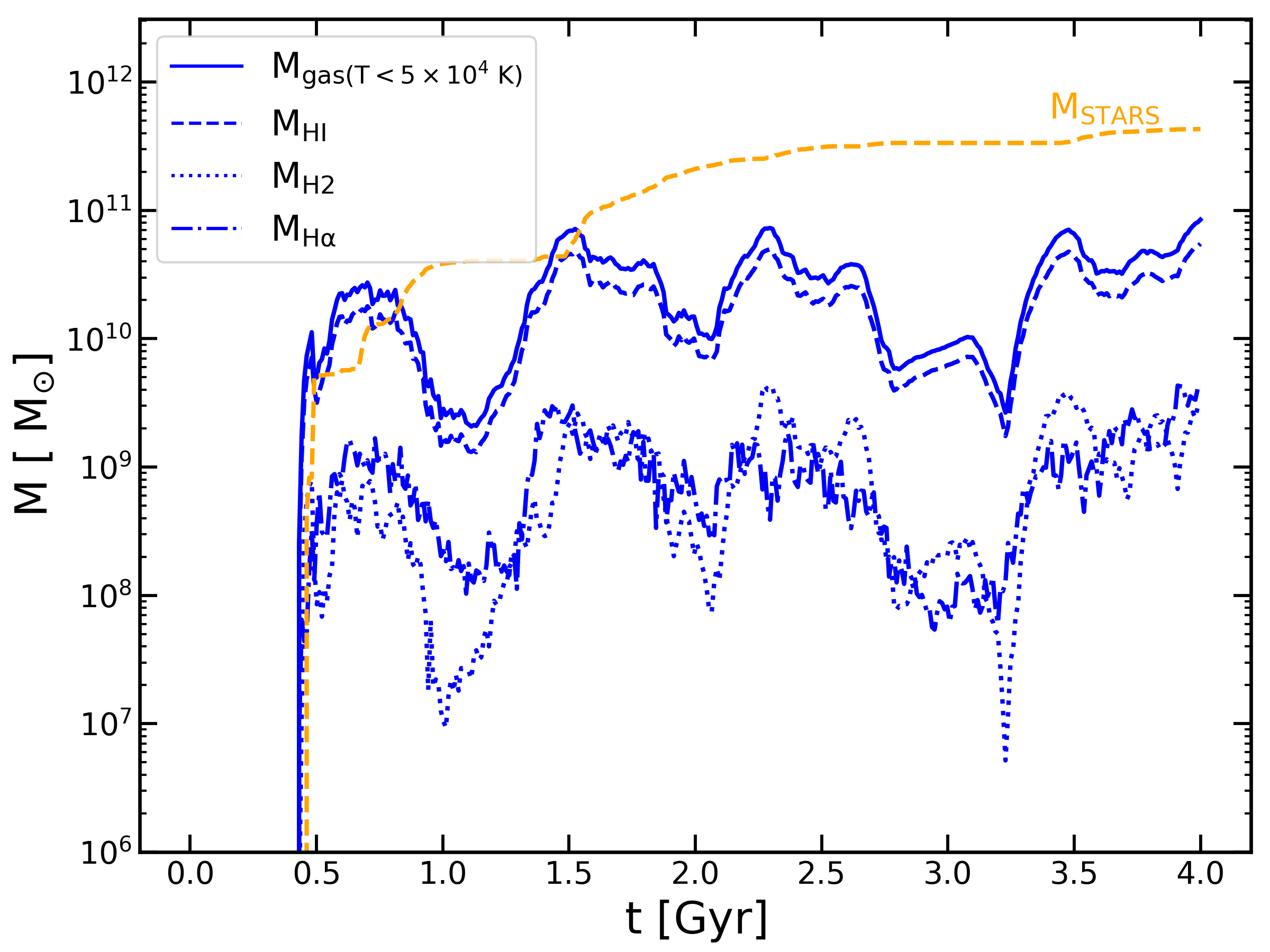}
\caption{Total mass of gas with temperature below $5\times10^4\ \mathrm{K}$ (solid blue line), as well as mass of neutral hydrogen $\mathrm{HI}$ (dashed blue line), molecular hydrogen $\mathrm{H2}$ (dotted blue line) and warm-ionized phase (dot-dashed blue line). Note that the total amount refers to both hydrogen and helium, while the different phases refer only to hydrogen. For completeness, the total stellar mass (orange dashed line) is also shown. All the quantities are largely consistent with the values observed in real systems. }
\label{fig:cold_mass_phases}
\end{center}    
\end{figure}

\begin{figure*}
\begin{center}
\includegraphics[scale=0.52]{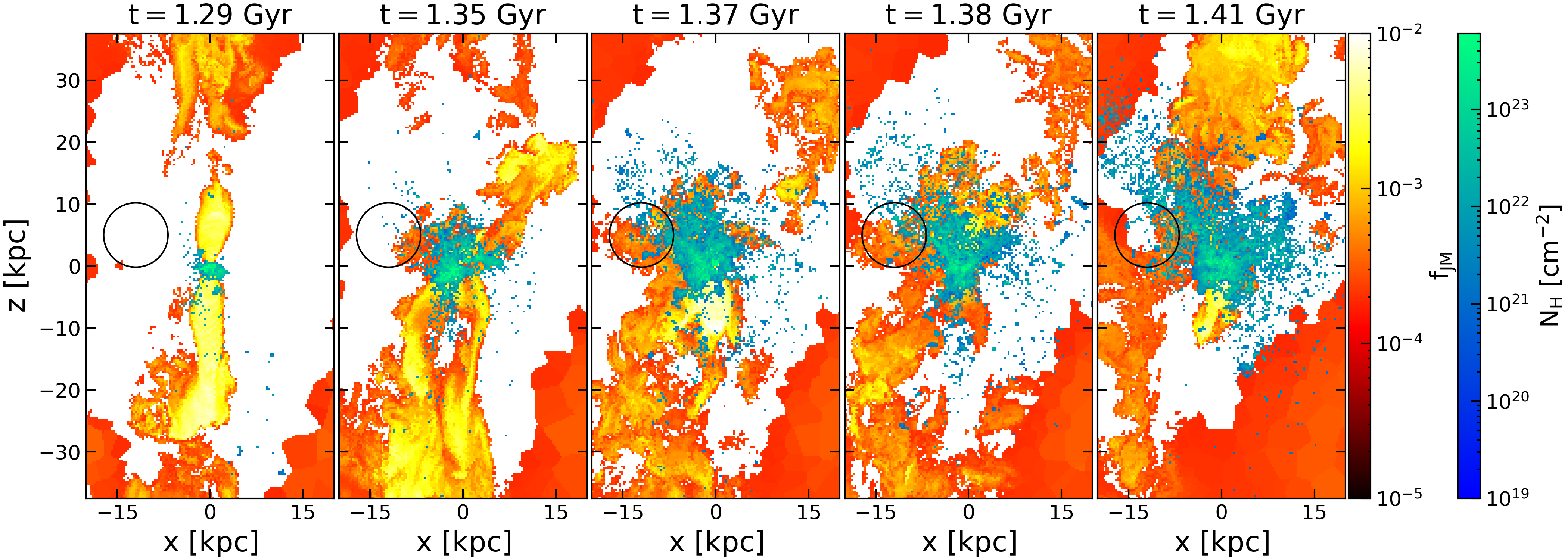}
\caption{A series of time instances showing the spatial distribution of the jet fraction in the meridional plane (red-yellow colour-coding). The column density of the gas with temperatures below $5\times10^4\  \mathrm{K}$ is overplotted as well (blue-green colour-coding).
The black circle indicates a $6\ \kpc$ radius region,  centred at $(x, y, z)=(-3, -12, -5)\ \mathrm{kpc}$, in which the properties of Fig.~\ref{fig:hist_hotGas_coolTime} are computed. As the jet material propagates laterally towards the indicated region it initially drags some of the cold gas with it ($t \sim 1.34-1.38$~Gyr); meanwhile, by injecting turbulence, creates the perturbations (Fig. \ref{fig:hist_hotGas_coolTime}) that, in the next $30\ \mathrm{Myr}$, form fresh cold gas in situ (Fig. \ref{fig:slice_initial_tracer}).A video of this figure is available in the online journal.}
\label{fig:snap_model}
\end{center}    
\end{figure*}

\begin{figure}
\begin{center}
\includegraphics[scale = 0.57]{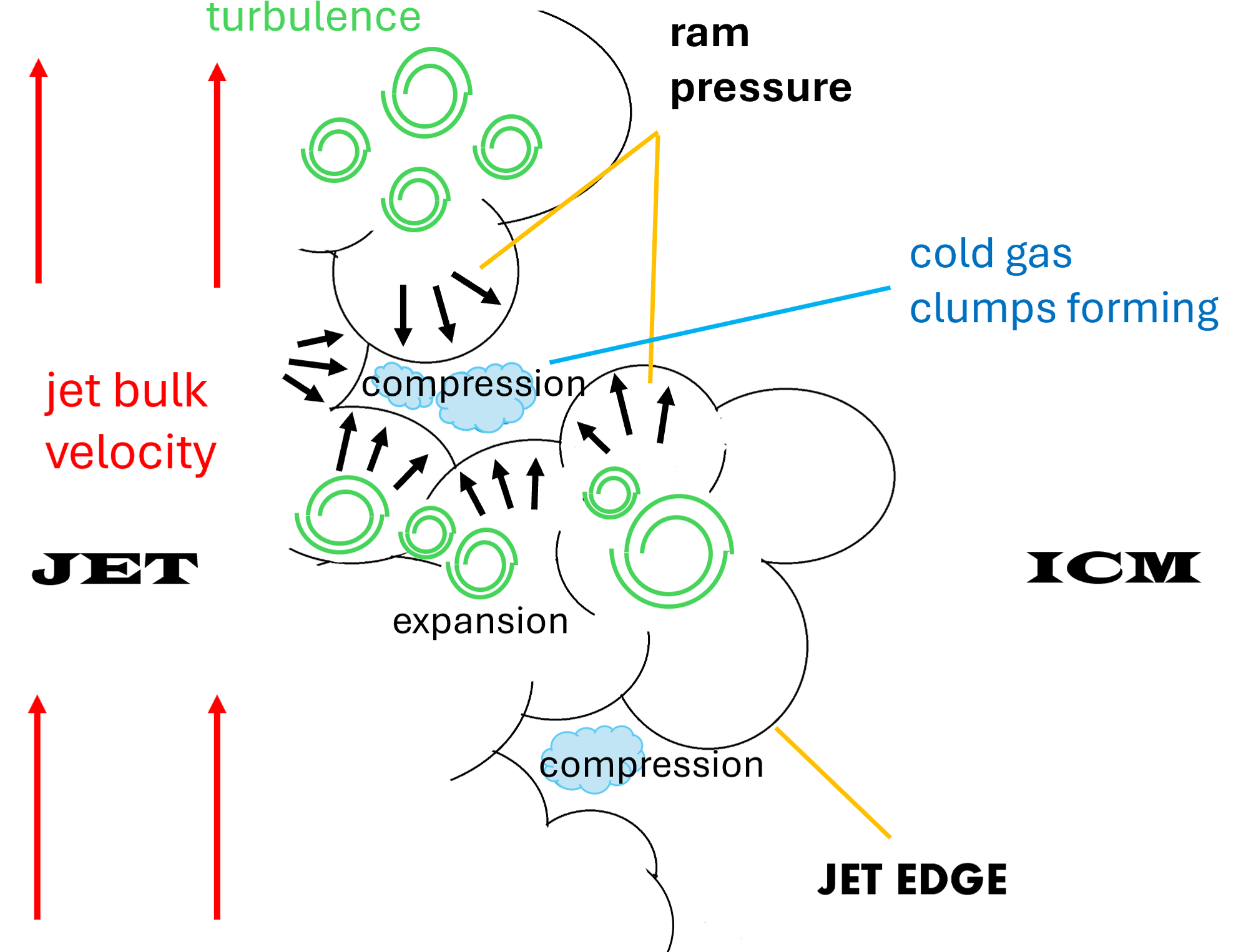}
\caption{Simplified sketch of the cold clumps formation model at the ``edges'' of the main jet. Only a small vertical section of one jet is shown for simplicity. The red arrow indicates the main velocity of the jet ($\sim 1000\ \kms$), while the edges are expanding in a perpendicular direction at a lower speed ($\sim 300\ \kms$).
The thin black line demarks the separation of the jet material from the ICM. 
The green curly regions indicate the presence of turbulence, and the bold black arrows indicate the pressure gradient direction. 
Turbulent vortices and black arrows marking pressure gradients are shown only in some parts of the sketch for clarity.
The blue clouds indicate the new cold clumps that are forming.
}\label{fig:sketch_model}
        \end{center}    
\end{figure}

During the entire simulation timespan of $4$~Gyrs, we identify several different stages of AGN feedback. In the first transient phase lasting $\sim 500\ \mathrm{Myr}$, influenced by the assumed initial conditions, the ICM is gradually cooling down, and as such, there is not enough cold gas to trigger the AGN activity. When only gravity and cooling processes are at play, the cold gas is mostly concentrated in the centre, as expected in classical cooling flow models, while, during, or shortly after AGN outbursts, it is distributed within larger and more extended structures. Both configurations are seen in observed galaxy clusters, and either one or the other can happen in the cluster centre, due to the long-term interaction with the AGN feedback (or lack thereof). The correlation, amount and morphologies of the cold gas with respect to the AGN power are not simple, as this interaction develops across a range of time and spatial scales.

When the AGN is activated by a sufficient quantity of cold gas reaching the accretion region, AGN power oscillates in the range $\sim 10^{44}-10^{45}\ \mathrm{erg\,s^{-1}}$ (bottom plot of Fig.~\ref{fig:visual}, grey curve). 
These values are consistent with observations \citep[e.g.,][]{Fabian12, 2012NJPh...14e5023M}. 
The AGN activity and the formation and evolution of cold gas in our simulation are two interconnected processes for the following reasons: a) as the cold gas forms in the ICM, a fraction of it falls onto the accretion region around the central SMBH, and ignites the AGN feedback; and b) when the AGN is on, its outbursts move a fraction of cold or cooling gas from the central region to larger radii, through uplift and entrainment, favouring further condensation \citep{Voit2017, Jennings2023}. Furthermore, as we will discuss later in greater detail, the AGN-driven jets create perturbations and compression regions in the hot ICM, leading to the formation of cold clumps and filaments \citep[see also][]{Gaspari18}. 

The amount of cold gas present in the ICM strongly correlates with the AGN power, due to both of the aforementioned effects.
Indeed, the evolution of the jets is very sensitive to the presence of cold gas in the central $\sim 10\ \kpc$ region. Specifically, we observe deflection and redirection of the jets, induced by interactions and collisions with dense cold clouds\footnote{We recall that in our adopted implementation, the SMBH has no natural precession and the jet axis is fixed.}\citep[see also][]{Ehlert23}. These deflections appear to be essential to distribute heating from jets, and to avoid the ``dentist drill'' effect: perfectly collimated jets, digging through the ICM, will travel a long path in the clusters and inject energy at large distances, as we found in \cite{Sotira25}, \citep[see also][]{VernReyn06, Libryan14b}. While in \citet{Sotira25}, precession of jets was introduced by the numerical setup; here, jet redirection is spontaneously induced by the interaction with the dense, multi-phase gas.  In the observed clusters, both the AGN jet precession, redirection via collisions with the dense clouds, and displacement by cluster weather \citep[e.g.,][]{2006MNRAS.373L..65H, BournSij19, bourneSij2021} are likely at play.

To verify that the multi-phase ICM properties are realistic, in Fig.~\ref{fig:cold_mass_phases}, we show the evolution of gas mass for the different phases of the hydrogen below $5\times 10^4 \ \mathrm{K}$, as well as the cumulative mass of the stars formed.
At the end of the simulation, the total stellar mass formed is $3\times10^{11}\ \msol$, $99\%$ being within $30\ \kpc$ radius, while the small remainder resides in filaments or clumps at larger radii. 
The total stellar mass, somewhat surprisingly, matches the cD galaxy of Perseus despite our simple stellar model; of course, in real systems, the central galaxy is not formed by a direct cooling flow as it happened in the initial stage of our simulation.
After the cooling flow stage, the average star formation rate is $\approx100\ \mathrm{\msol/yr}$. This value would mostly produce a blue star-forming central galaxy rather than a massive red passive galaxy as observed. Several key aspects that can suppress the star formation are missing in the simulation, such as most importantly stellar feedback. Furthermore, our simulations assume a simple prescription for star formation, while the injection of turbulence from the AGN itself and the presence of a magnetic field can drastically reduce the star formation efficiency of the molecular phase \citep{Alatalo2015, Tabatabaei2018}. 
We again stress the simple nature of the star formation model employed in our simulations, which is, at least partly, employed for numerical reasons, to avoid excessively dense gas formation. 
Importantly, the realistic stellar mass in conjunction with the reasonable black hole and cold gas mass we find in our simulation demonstrates the self-regulated nature of our AGN feedback implementation, where the numerical overcooling is avoided. The SMBH mass is quite constant during the whole simulation, growing by roughly $10\%$ from its initial mass of $M_{\mathrm{BH}}=4\times10^9\ \msol$.

Approximately $90\%$ of the warm-cold gas is neutral, while the remainder is in molecular form, with masses in the range $10^7-2\times10^9 \ \msol$. The mass of the warm-ionized hydrogen component is of the same order of magnitude as the molecular gas. These values are compatible with observations of observed systems \cite{Salome11, Olivares2019, olivares22, Gingras24}, except for the warm-ionized gas, which appears to be higher in our model with respect to observations by 
\citet{Gingras24}, who found ionized gas masses in the range $10^6 - 10^8$ M$_\odot$ in a small sample of cool-core clusters.

\subsection{Cold gas formation model}\label{sec:cold_gas_formation}

\begin{figure*}
\begin{center}
\includegraphics[scale =0.48]{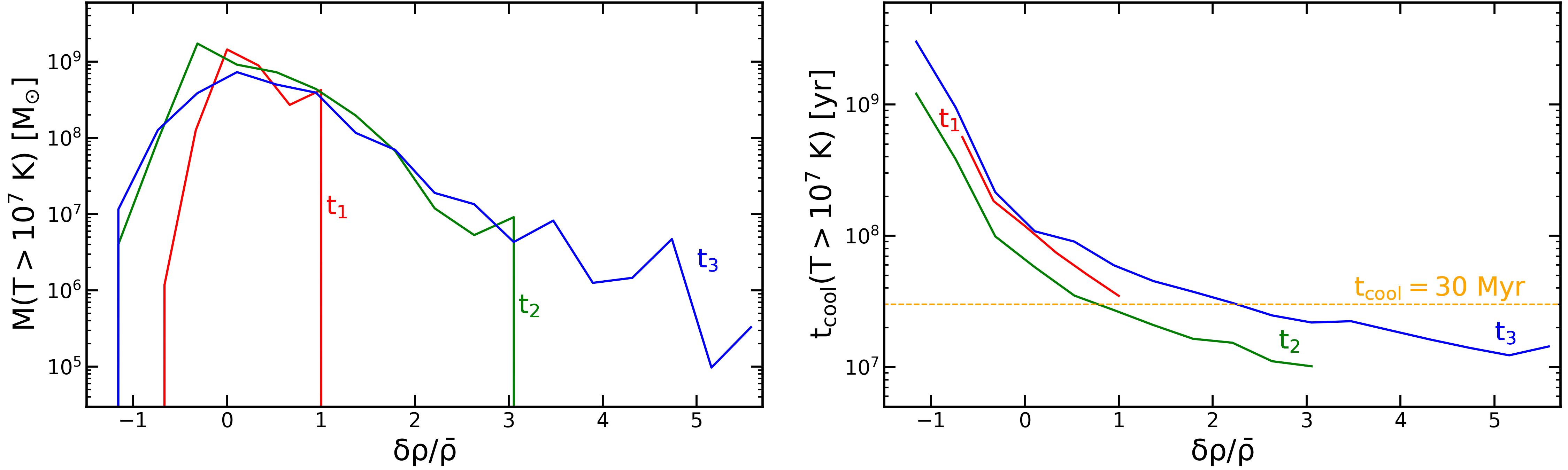}
\caption{Left: Distribution of the hot gas mass, i.e., with $T>10^7\ \mathrm{K}$, binned as a function of overdensity, $\delta\rho/\bar{\rho} = (\rho_i-\bar{\rho})/\bar{\rho}$, where $\rho_i$ and $\bar{\rho}$ are the density of the $i$-th cell and the average density, respectively. Right: corresponding mass-weighted cooling time in each bin of overdensity. 
The distributions are computed in a spherical region having a radius of $6\ \kpc$ and centred at $(x, y, z)=(-3, -12, 5)\ \mathrm{kpc}$ (highlighted by the black circle in Figs.~\ref{fig:snap_model} and \ref{fig:slice_initial_tracer})  at $t_1 = 1.29\ \Gyr$ before the jet passage (red line), $t_2 = 1.35\ \Gyr$ and $t_3=1.38\ \Gyr$ during the jet passage (green and blue lines). While the distribution of overdensities is very narrow at $t_1$, at $t_2$ and $t_3$ there is a broad tail of much larger overdensities.
The orange dashed line in the right panel separates bins that have average cooling times below $30\ \mathrm{Myr}$. After the passage of the AGN jet, a significant fraction of the hot ICM mass exhibits low cooling times.}
\label{fig:hist_hotGas_coolTime}
\end{center}    
\end{figure*}

\begin{figure}
\begin{center}
\includegraphics[scale =0.45]{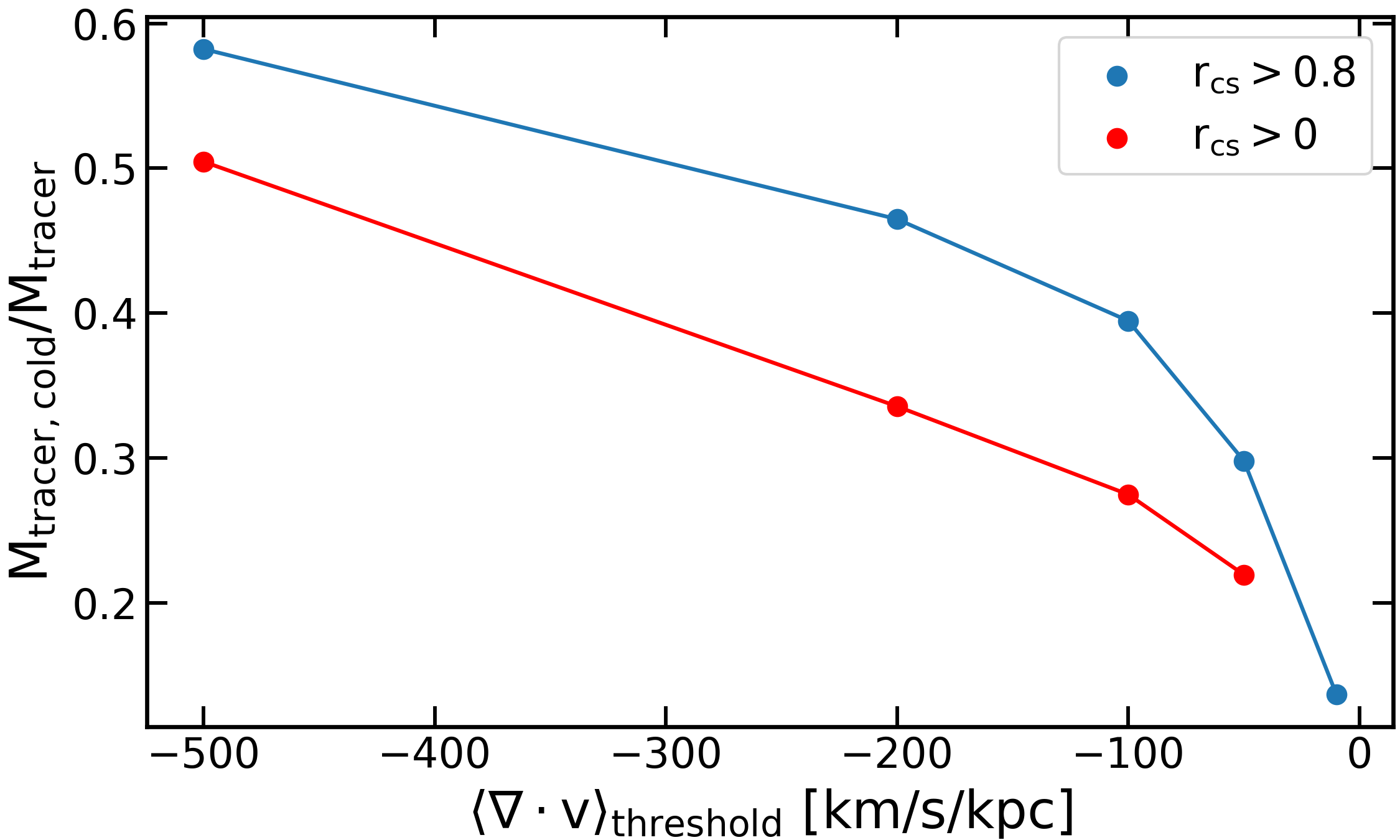}
\caption{Mass fraction of tracers that goes in the cold phase for every tracer re-run (dot symbols) performed with different divergence cuts from $-10\ \kms/\mathrm{kpc}$ to $-500\ \kms/\mathrm{kpc}$ ($x$-axis) and with different compressive ratio cuts, $r_{\mathrm{cs}}>0.8$ (blue line) and $r_{\mathrm{cs}}>0$ (red line). The mass of tracers that condense in the cold gas increases when the divergence and compressive ratio cuts are higher, reaching up to $\sim 60 \%$.}
\label{fig:frac_runs_tracer}
        \end{center}    
\end{figure}

In this section, we quantitatively investigate the formation scenario of cold clumps. Once non-linear density perturbations are generated, they can undergo rapid localized cooling. 
Such perturbations preferentially originate in regions of sustained compression in the ICM, due to bulk motions and turbulence, naturally driven by the central AGN-driven outflows 
\citep{Revaz08, BrigMath02, Gaspari13, Gaspari18, Brigh15, LiB15, Sotira25}.

In Fig.~\ref{fig:snap_model}, we show central slices of the jet mass fraction, i.e., the mass added to the launching cells in the injection cylinder over the total mass of the cells, which is displayed by the yellow-red colour tones\footnote{The fraction of the jet mass is low, because this mass is added to cells already present in the injection cylinder, i.e., it is only the $dm$ in Eq.~\ref{eq:mass_injection}{\color{red}{b}}.}, at different times. 
The yellow tone mostly marks material from the newly injected jet, while redder tones mark older and more diluted jet material\footnote{For clarity, we do not display here jet material with a fraction less than $10^{-4}$.}. Notice the presence of the jet redirection due to encounters with cold clumps.

Fig.~\ref{fig:snap_model} also shows the column density of the gas with temperatures below $5\times10^4\ \mathrm{K}$ (blue-green colour tones). 
The cold clumps are found displaced from the jet main axis, with some of them spatially located within the laterally expanding jet lobes edges. 
The displacement is due to the fact that, soon after an AGN jet outburst, small amounts of cold gas follow the jets as a result of entrainment and ram pressure \citep{Voit2017}; other condensation can be further enhanced during fragmentation of the cold clouds by the hot jets \citep{Jennings2023, Ghosh2025}. 
Another mechanism proposed is the cold cloud formation in situ due to the AGN induced turbulence \citep[e.g.;][]{Gaspari18}; we explore these ideas further by linking the formation of cold clumps to the turbulent expansion and compression of the jet, especially when the jets expand or are re-directed toward unperturbed regions of the ICM.

\begin{figure*}
\begin{center}
\includegraphics[scale =0.455]{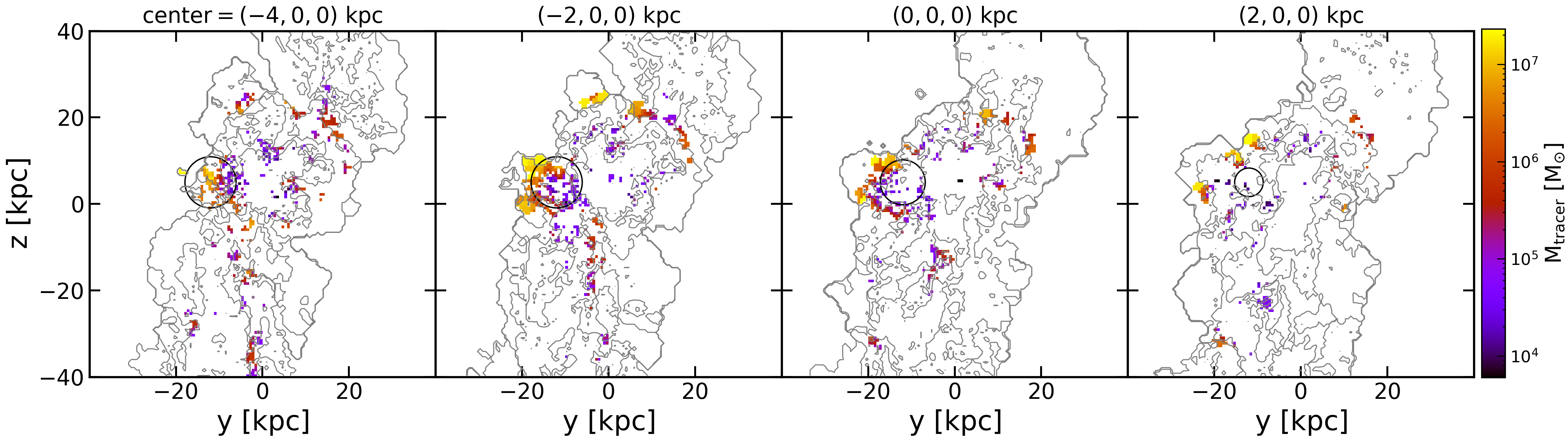}
\caption{Different slices in the $y-z$ plane of the jet contours (grey) and tracer mass map at the time of the injection using the following criteria: $\langle\nabla\cdot v\rangle<-100\ \kms/\kpc$ and $r_{\mathrm{cs}}>0.8$ (run $A$), centred at the cluster centre, and $\pm2$, and $4\ \kpc$ away from the centre, from left to the right, respectively. The tracers are placed where the divergence is strongly negative (i.e., zones of compression) and where the compressive ratio is high (i.e., zones with low vorticity). The maps provide a simulation counterpart to the scenario explained in the sketch of Fig.~\ref{fig:sketch_model}.
The black circle indicates the intersection of the spherical region of Figs. \ref{fig:snap_model} and \ref{fig:hist_hotGas_coolTime} with the different slices. The presence of tracers within it confirms the presence of perturbations that will condense in cold clumps. A video showing the relation between jets, tracers and cold gas is available in the online journal.}
\label{fig:slice_initial_tracer}
        \end{center}    
\end{figure*}

As a representative example of this mechanism, we consider a spherical region centred at $(x,y,z)=(-3,-12,5)\ \kpc$, and having a radius equal to $6\ \kpc$ (highlighted by the black circle in Figs.~\ref{fig:snap_model}) which is unperturbed at $t=1.29\ \Gyr$ and has small turbulence with velocity dispersion\footnote{With $\sigma_{\mathrm{hot}}^2=1/3(\sigma_x^2+\sigma_y^2+\sigma_z^2)$, where the terms on the right-hand side of the equation are mass-weighted velocity dispersions along each axis.}, $\sigma_{\mathrm{hot}}$, of the hot phase around $30\ \kms\approx0.05\ c_{s,\mathrm{hot}}$, where $c_{s,\mathrm{hot}}$ is the mass-weighted sound speed of the hot phase in the region (see also Section~ \ref{SubSec:turbulence}).
At $t = 1.35\ \Gyr$, part of the jet bends while hitting the cold gas in the cluster centre, and propagates in a perpendicular direction ($x$) with respect to the jet axis ($z$); in the time span $t = 1.35-1.38\ \Gyr$, the jet material enters the region injecting turbulence and creating the perturbations for the cold clumps that condense in the next $30\ \mathrm{Myr}$, as we will show in Section~\ref{subsec:tracer} (Fig.~\ref{fig:slice_initial_tracer}).
At $t=1.38\ \Gyr$, the velocity dispersion of the hot phase in the region is $\sigma_{\mathrm{hot}}=120\ \kms\approx0.2\ c_{s,\mathrm{hot}}$.

The sketch in Fig.~\ref{fig:sketch_model} represents a schematic view of the likely formation scenario for the cold clumps in our simulation, which we will investigate quantitatively in the rest of this Section. We find that at a given time, a jet cocoon can expand laterally with relatively low velocities, $300-500 \kms$, with respect to the jet main axis (indicated by the red arrows), and move towards unperturbed portions of the ICM.
The green curly regions denote the presence of turbulence, and the bold black arrows indicate the ram pressure (both are shown only in some parts of the sketch for clarity).
In this configuration, the turbulent expansion triggers density enhancements in the ICM, which promote cooling into dense clumps. 
At later times, these structures can merge into larger clouds or filaments and, unless they get heated by an AGN outburst, they will migrate towards the cluster centre \citep[see also][]{Gaspari13}. 
At the front of the jet, there is instead no cold gas condensation as the jet moves supersonically, and even though there is compression, it is also associated with strong shock heating. So even in the case that gas clumps are present ahead of the expanding jet, their fate is to ultimately be disrupted and heated by the violent jet activity there.

\begin{figure}
\begin{center}
\includegraphics[scale =0.49]{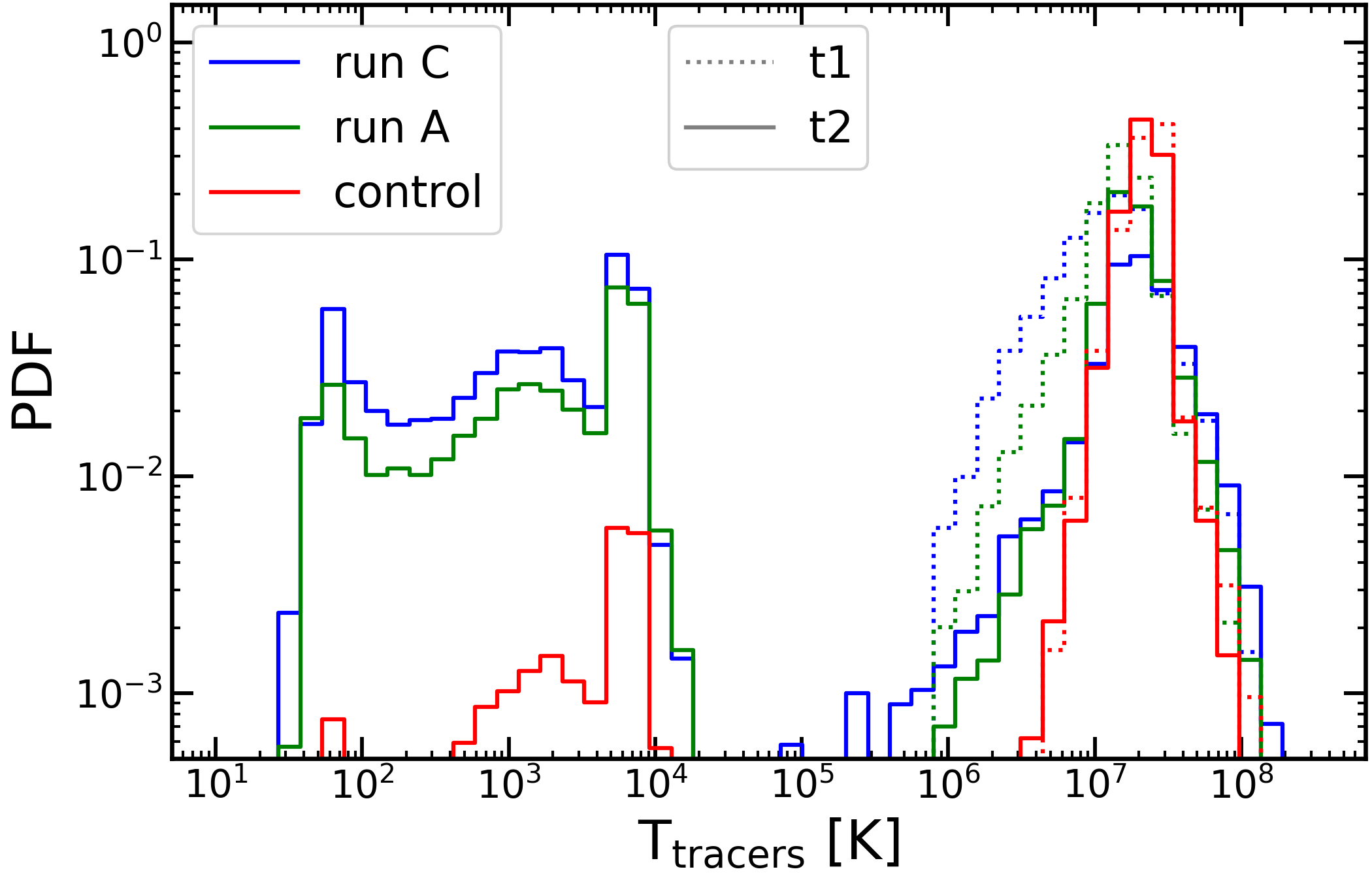}
\caption{Temperature PDF of the tracers at time of the injection (dotted lines) and after $30\ \mathrm{Myr}$ (solid lines), for runs $A$ (green), $C$ (blue), and control (red), corresponding to
$\langle\nabla\cdot v\rangle$ lower than $-100$, $-500\ \kms/\kpc$ and $r_{\mathrm{cs}}>0.8$ for runs $A$ and $C$, and $\langle\nabla\cdot v\rangle>0$ without the $r_{cs}$ cut for the control run. Our conditions on velocity divergence and compressive ratio successfully identify regions where cold clumps form in situ from the hot gas.}
\label{fig:histogram_tem_tracer}
        \end{center}    
\end{figure}

The key physical driver of the cooling process is the increase in the gas velocity dispersion, which, combined with the particular geometry of the jet expansion, creates multiple zones of high compression, leading to thermally unstable portions of hot gas. This is quantitatively shown in Fig.~\ref{fig:hist_hotGas_coolTime} for the spherical region described above (highlighted by the black circle in Fig.~\ref{fig:snap_model}). 
The left-hand panel shows the distribution of the hot gas mass ($T>10^7\ \mathrm{K}$) with respect to the local overdensity. Here $\delta\rho/\bar{\rho}= (\rho_i-\bar{\rho})/\bar{\rho}$, where $\rho_i$ and $\bar{\rho}$ are the density of the $i$-th cell and the average density (in the spherical region), respectively. 
The right-hand panel of Fig.~\ref{fig:hist_hotGas_coolTime} instead shows the mass-weighted average cooling time in each overdensity bin \footnote{Note that we excluded from this analysis gas with temperature ranging between $\approx10^6$ and $\geq 10^4\ \mathrm{K}$, as it already has a short cooling time and a significant fraction of it will soon become a warm ionized medium with $T\approx10^4\ \mathrm{K}$.}.
The distributions in Fig.~\ref{fig:hist_hotGas_coolTime} are shown at three different times, before and during the jet passage.
At $t_1$, just before the passage of the redirected jet burst, the hot gas mass distribution (left-hand panel; red curve) is approximately symmetric and centred around $\delta\rho/\bar{\rho}\sim 0$, with gas cooling time never below $30\ \mathrm{Myr}$ (right-hand panel).
After the jet passage, at $t_2$ and $t_3$, the mass distribution has significantly changed (left-hand panel; green and blue curves); the distributions are now skewed towards densities much larger than the average, with large amounts of hot gas having $\delta\rho/\bar{\rho}=3-5$, indicating significant gas compression. Some of this compressed gas has a cooling time ranging in $10 - 30\ \mathrm{Myr}$ (right-hand panel; green and blue curves). 

In our simulation, such strongly non-linear overdensities are frequently formed by propagating jet material, and they eventually cool down to form the cold gas later found in the same region, before falling towards the centre. 
This happens despite the fact that the majority of the hot gas there has instead increased its cooling time: this means that the passage of the jet heats up overall the region, both via mixing with hot jet material, (weak) shocks and turbulent dissipation, while at the same time the same process promotes the localized formation of a significant amount of cold gas. 

\subsection{Tracking cold gas formation with tracers}\label{subsec:tracer}
To further explore the validity of this model, we re-ran the simulation introducing a passive tracer fluid into gas cells satisfying different physical criteria. This allows us to determine the fraction of hot gas that, at any time in the simulation, is most likely to undergo the formation of cold clumps at later times.
In particular, we re-run the simulation in the time span $t=1.38 - 1.5\ \Gyr$, with a maximum time step of $2\  \mathrm{Myr}$ to follow the cooling processes in detail. 
We kept track of regions experiencing significant compression by introducing tracers based on specific ranges of values of velocity divergence and of compressive ratio at the injection epoch. 
The latter is computed following \citet{Iapichino2011,BourneSij17}, as 
\begin{equation}
    r_{cs} = \frac{\langle\nabla\cdot v\rangle^2}{\langle\nabla\cdot v\rangle^2+\langle\nabla\times v\rangle^2},
\end{equation}
where the average $\langle\rangle$ is computed mass-weighting over 32 neighbouring cells.
We ran a large number of simulations by considering gas cells that have a mass-weighted divergence, $\langle\nabla\cdot v\rangle$, lower than a series of values ranging from $-500$ to $-10\ \kms/\kpc$ with (and without) $r_{\mathrm{cs}}>0.8$ to keep track of the jet compressive flow \citep{BourneSij17}.
We additionally ran two control simulations with $\langle\nabla\cdot v\rangle\geq 0$ and $>+100 \kms/\kpc$, corresponding to barely or strongly expanding regions.
For all the runs, the selection criteria, the initial selected mass tracked by the tracer fluid and the percentage of tracer mass that goes in the cold phase are listed in Table~\ref{tab:tracer_runs} in the Appendix. 

In the selection process of the gas cells, a first selection considered cells with temperatures higher than $10^7\ \mathrm{K}$. Their 32 neighbouring cells are then used to compute the average divergence and compression ratio, which we filter on to localize compression zones. 
However, since by considering neighbouring cells gas with a range of temperatures may be included, we additionally remove cells with $T<10^6\ \mathrm{K}$ from our tracer injection region. We note that, after this cut, cells with temperature between $10^6\ \mathrm{K}<T<10^7\ \mathrm{K}$ contribute only a small fraction of the total mass.
The selected mass in which tracer are injected ranges from $2$ to $8\times10^9\ \msol$ for runs $A$, $B$, $C$, and is $2.8\times10^{11}$ for the control run.

Fig.~\ref{fig:frac_runs_tracer} summarizes the main results of these tests and gives the mass fraction sampled by tracers which has condensed into cold ($ \leq 5\times10^4\ \mathrm{K}$) gas after 30~Myr, as a function of the value of velocity divergence chosen for the initial selection. 
The blue and the red lines in the plot indicate runs with and without the compressive ratio cut, $r_{\mathrm{cs}}>0.8$, respectively. 
We can see a clear trend: the mass fraction of tracers that condense into warm and cold phases is higher when increasingly more negative values of divergence are used for the initial selection of tracers.
Furthermore, the compressive ratio also plays a role since the fraction is always higher when the $r_{\mathrm{cs}}$ threshold is applied. 
In other words, the level of ongoing compression in the hot gas phase at a given epoch appears to be a good predictor of gas undergoing future condensation into cold clumps, after $\sim 30\ \rm Myr$. 

While this trend can qualitatively be predicted a priori, we show that our selection criterion specifically determines cells that are affected by the jet perturbation and therefore produces a tangible {\it positive} feedback promoting cold gas formation. 
Let us  consider our fiducial runs with $r_{\mathrm{cs}}>0.8$ and  $\langle\nabla\cdot v\rangle$ lower than $-100$ (run $A$) and $-500\ \kms/\kpc$ (run $C$), corresponding respectively to a compression time, $t_{\mathrm{comp}}= \langle\nabla \cdot \textbf{v}_i\rangle^{-1}$, of $10$ and $3\ \mathrm{Myr}$.
Fig.~\ref{fig:slice_initial_tracer} shows different slices of the injected tracers at the initial time for run $A$, together with the jet contour in grey.
The tracer presence indicates zones with high negative divergence and relatively low vorticity (high compressive ratio).
This selection marks regions highly perturbed by the jet, and the injected tracers are confined to localized regions at the edges of tangled jet contours.
The location of such selected regions, i.e., at the tangled interface between laterally expanding jets and the ICM, qualitatively confirms the model already introduced in the sketch of Fig.~\ref{fig:sketch_model}.
Furthermore, the black circle in the figure indicates the region analysed in the previous Section (Figs. \ref{fig:hist_hotGas_coolTime} and \ref{fig:slice_initial_tracer}). We note that at $t=1.38\ \Gyr$, a large amount of tracer mass is present in the region, confirming what is pictured in Section~\ref{sec:cold_gas_formation}. 

Finally, Fig.~\ref{fig:histogram_tem_tracer} shows the temperature distribution functions for run $A$ (green lines), $C$ (blue lines), and control (red lines) at the moment of the injection (dotted lines) and the time of the peak of tracers in the warm and cold phases i.e. $30\  \mathrm{Myr}$ later (solid lines).
In runs $A$ and $C$, the initial distribution of hot gas associated with tracers has been roughly halved after $30\ \mathrm{Myr}$, with the other half of the initially selected mass being now in the cold phase. On the other hand, 
the initial temperature distribution of the control run remains almost the same, only with a small fraction of gas ($\leq 10 \%$) collapsing into the cold phase. 

\begin{figure}
\begin{center}
\includegraphics[scale =0.45]{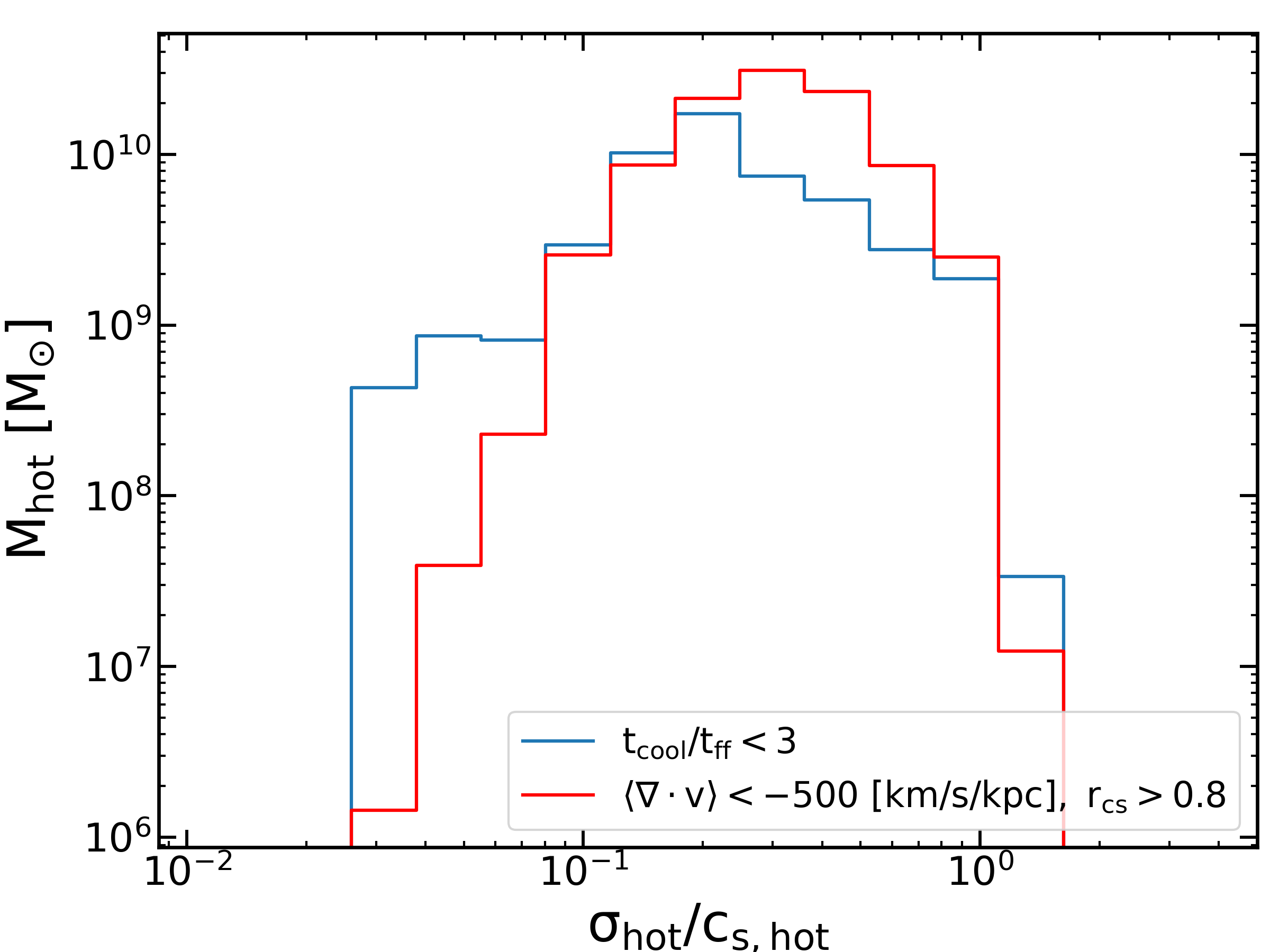}
\caption{Hot gas mass, i.e. with temperatures higher than $T>10^7\ \mathrm{K}$, that has $\langle\nabla\cdot v\rangle<-500\ \kms/\kpc$ and $r_{\mathrm{cs}}>0.8$ (red line) and $t_{\mathrm{cool}}/t_{\mathrm{ff}}<3$ (blue line) computed for a sample of $5\ \kpc$ radius regions in the cluster core for $t = 1.3-2.5$~Gyr, computed against the total velocity dispersion of each region normalized to its mass-weighted average sound speed.}
\label{fig:massFraction_inst_cool}
        \end{center}    
\end{figure}

\subsection{Link to the ICM turbulence}\label{SubSec:turbulence}

The cold clumps formation scenario singled out by our simulation implies a tight relation with the turbulent motions episodically stirred by jets expanding in the ICM. Therefore, it is of interest to link the presence of positive AGN feedback with the macroscopic observables that are now potentially detectable via X-ray spectroscopy by the XRISM satellite \citep[e.g.][]{Fujita25}.
To statistically study this, we sampled the central cluster volume by selecting 7 spherical regions with a radius of $5\ \kpc$. 
The regions are centred on $(0,0,\pm10)$, $ (\pm10,0,0)$, $(0,\pm10,0)$ and $(0,0,0)$, i.e the cluster centre. 
In the time span $1.3-2.5\ \Gyr$, we sum up the total gas mass of cells with temperatures $\geq 10^7$~K that have  $\langle\nabla\cdot v\rangle<-500\ \kms/\kpc$ and $r_{\mathrm{cs}}>0.8$ (as done in our tracer analysis) and plot the resulting mass as a function of the turbulent Mach number of the corresponding region at that moment in Fig.~\ref{fig:massFraction_inst_cool} (red line). 
The latter is calculated as the total gas velocity dispersion over the mass-weighted sound speed in the region, $c_{s,\mathrm{hot}}$, at that moment. The turbulent Mach number is computed only for gas cells with temperatures $\geq 10^7\ \mathrm{K}$.

The resulting gas mass distribution has roughly a Gaussian shape, peaked at $\sigma_{\mathrm{hot}}\approx0.3\ c_{s, \mathrm{hot}}$.
We compute, as well, for each region the hot gas mass of cells which have free-fall times, computed as $t_{\mathrm{ff}}=(2r^2/g)^{1/2}$ (where $r$ is the distance of the region from the centre and $g=g(r)$ is the gravitational acceleration at that distance) shorter than three times the cooling time, i.e. $t_{\mathrm{ff}}/t_{\mathrm{cool}}<3$ (blue line). 
Hot gas cells with the latter ratio of the order of unity are bound to condense and form cold clumps, before they can reach the cluster centre \citep{Malagoli87, Loewenstein1999,Voit2021}.
The two distributions in Fig.~\ref{fig:massFraction_inst_cool} are roughly the same, which is in line with the expectation of the cold clumps formation scenario we outlined above. 

This result indicates that values of $\sigma_{\mathrm{hot}}$ around $1/3$ of the sound speed are necessary to promote cold gas condensation (Fig.~\ref{fig:sketch_model}), see also \citet{Gaspari18}.
The condition on $\sigma_{\mathrm{hot}}$ must be interpreted as a necessary but not a sufficient condition to create cold clump condensation. Indeed, there are several periods of time during which regions have those values of turbulent Mach number, but not cells with the criteria for condensation.
This can happen, for example, when the cold gas condensation has already occurred and the region maintains the same value of turbulence, but the gas is predominantly too hot.

Cells which are subject to strong compression also are associated with high local values of the velocity dispersion in the hot gas, and will be forming cold clumps given their short cooling time at a given radius; only the passage of AGN jets can promote such high values of velocity dispersion in the hot gas in our simulation, and this once more proves the positive feedback loop between jets, turbulence, compression and cooling in this numerical model.
In principle, using  X-ray spectroscopy like the ones provided by XRISM \citep{XRISMColl25may, XrismCol25March}, future observations should be able to identify the most likely regions of the ICM where cold gas clumps should undergo formation, promoted by the interaction with AGN jets. Spatially co-locating these regions with deep, high-resolution radio maps may provide further evidence that lateral jet material expansion drives localised gas compression and in-situ cold clump formation.

\section{Discussion and conclusions} \label{sec:conclusions}

The idea that feedback from the central AGN enhances off-centre gas condensation in the ICM, and other astrophysical environments, has been often proposed in the past decades \citep{gaspari12, Gaspari18, Voit2017}. In this work, we used advanced numerical simulations of high-resolution AGN jets in an isolated Perseus-like cluster to better investigate this phenomenon. Our model captures a self-consistent SMBH accretion and feedback cycle in a fully multi-phase ICM medium, ranging from the hot, X-ray emitting to the molecular gas, whereby the mass in different gas phases, ICM radial profiles and AGN power are in very good agreement with observations.

We have studied a specific mechanism where jet-induced off-axis turbulent motions in the hot gas lead to compression and in-situ condensation into cold clumps, typically on a time scale of a few $10\mathrm{s}\ \mathrm{Myr}$. 
The proposed process happens simultaneously while the jet heats up the ICM and increases its entropy to the levels consistent with the cool-core clusters. 
We assessed the validity of the proposed cold gas formation model by using tracers, which allowed us to directly link in the simulation localised compression regions in the hot gas (spatially associated with jet propagation sites) and the formation of cold clumps from the same perturbed gas after about a cooling timescale. 

After having highlighted and tested the validity of the proposed model, we want to stress that this scenario is not the only channel to produce cold gas in cluster cores. Indeed, the entire amount of warm and cold gas found outside of the innermost regions is not only generated starting from jet-induced compression, as it is unlikely that this mechanism alone has a sufficient efficiency for this task. 
For example, also the advection and uplift of cold or low entropy gas from the centre, due to the jet expansion, has been proposed to promote further cold clump formation, while entraining and mixing with the hotter and less dense gas outside of the cluster core \citep{gaspari12, Voit2017, Jennings2023}. 
Furthermore, in observed galaxy clusters, density fluctuations that lead to thermal instabilities can be generated by other astrophysical processes like galaxy wakes, cosmological accretion, and compressive turbulence driven by other sources \citep{Voit2018, Voit2021, Choudhury2019, Fielding2020}. 

Our model is not in contrast with this global picture, but it highlights a further important effect, which we were able to isolate thanks to the very high spatial resolution of jet lobes in our simulation and detailed gas radiative cooling modelling down to the molecular phase.
To summarize, our main conclusions are as follows:\\
$\bullet$ The compressive motion induced by the (lateral) jet expansion strongly widens the overdensity distribution and generates relatively dense localized regions in the ICM, where cold gas clumps form on a time scale of $\sim 30\ \mathrm{Myr}$;\\
$\bullet$ The regions of the hot ICM that experience this rapid cooling process are those characterised by significant compression and low absolute value of vorticity; \\
$\bullet$ Statistically, this process prevalently occurs when the local turbulent Mach number in the hot gas is $\sigma_{\mathrm{hot}} / c_{s, \mathrm{hot}} \approx 0.3$;\\
$\bullet$ The process of cold clump condensation (positive feedback) happens despite the presence of the AGN-driven jet distributed heating (negative feedback), which prevents the over-cooling catastrophe.

The formation process for cold gas clumps identified in this work is physically well-motivated; although only future, more systematic studies will allow us to quantify its efficiency in converting hot into cold gas, as well as to generalise its validity to a larger number of host systems. 
However, integrated with a model which can explain how hot, rarefied jets can uplift cold gas from the centre to large distances, this study sets the basis for a more complete scenario that incorporates key physical mechanisms for cold gas formation and redistribution. Future observational data, including X-ray spectroscopy, molecular and warm gas kinematics, as well as deep, high-resolution jet lobe radio maps, in conjunction with detailed AGN-driven jet simulations, such as the ones presented here, will finally help us to unlock the nature of the puzzling filamentary nebulae residing in the centres of cool-core clusters.

\begin{acknowledgements}
MAB is supported by a UKRI Stephen Hawking Fellowship (EP/X04257X/1). DS acknowledges support from the Science and Technology Facilities Council (STFC) under grant ST/W000997/1.
FV has been partially supported by Fondazione Cariplo and Fondazione CDP, through grant n$^\circ$ Rif: 2022-2088 CUP J33C22004310003 for the "BREAKTHRU" project, and by the European Union’s  Horizon Europe program through the ERC Synergy Grant COSMOMAG (Project Id. 101224803). 
This work used the DiRAC Data Intensive service (CSD3 [*]) at the University of Cambridge, managed by the University of Cambridge University Information Services on behalf of the STFC DiRAC HPC Facility (\url{www.dirac.ac.uk}). The DiRAC component of CSD3 at Cambridge was funded by BEIS, UKRI and STFC capital funding and STFC operations grants. DiRAC is part of the UKRI Digital Research Infrastructure.

All the text and the codes for the plots and simulation analysis throughout this paper have been written by humans.
\end{acknowledgements}

\bibliographystyle{aa}
\bibliography{biblio_Ste}

\onecolumn
\appendix
\section{Table of the re-runs with tracers}
\begin{table*}[h]
    \centering
    \begin{tabular}{cccccccccccc}
    \hline & run $Y$ & run $Z$
 & run $A$ & run $B$ & run $ C $& run $D$ & run $E$ & run $F$ & run $G$ & control & control2 \\
    \hline
$r_{cs}$ & $>0.8$ & $>0.8$ & $>0.8$ & $>0.8$ & $>0.8$ & - & - & - & - & - & - \\
$\langle\nabla\cdot v\rangle$ &$\leq-10$ &$\leq-50$ &$\leq-100$ & $\leq-200$  & $\leq-500$ & $\leq-50$ &$\leq-100$ & $\leq-200$  & $\leq-500$ & $>0$ & $\geq+100$  
\\ 
$M_{\mathrm{in}}\  [10^9\ \msol]$ & $51$ & $15.5$
& $8$ & $5$ & $2$ & $28.4$ & $15.9$ & $8$ & $2.4$ & $279$ & $16$ \\

$M_{\mathrm{fin,cold}}$& $13.7\%$ &$29.8\%$
& $39.5\%$ & $46.5\%$ & $58.2\%$ & $21.9\%$ & $27.5\%$ & $33.6\%$ & $50.5\%$  & $13.6\%$ & $12.2\%$ \\
    \hline
    \end{tabular}
    \caption{First and second rows: compressive ratio and divergence criteria used to inject the tracers.
    Third row: total gas mass selected using the above criteria.
    Fourth row:
    percentage of tracer mass that condenses in warm and cold clumps ($T<5\times10^4\ \mathrm{K}$) with respect to the total mass of tracer injected, for the various re-runs.}
    \label{tab:tracer_runs}
\end{table*}

\end{document}